\documentclass[aps, prb, twocolumn,amsmath,amssymb,floatfix, reprint, longbibliography]{revtex4-2}
\usepackage{graphicx}
\usepackage{physics}
\usepackage{times}
\usepackage{dcolumn}
\usepackage{hyperref} 
\hypersetup{colorlinks, allcolors=blue}

\begin{document}

\title{Quantum topological phase transitions in skyrmion crystals}

\author{Kristian M{\ae}land} 
\affiliation{\mbox{Center for Quantum Spintronics, Department of Physics, Norwegian University of Science and Technology, NO-7491 Trondheim, Norway}} 
\author{Asle Sudb{\o}}
\email[Corresponding author: ]{asle.sudbo@ntnu.no}
\affiliation{\mbox{Center for Quantum Spintronics, Department of Physics, Norwegian University of Science and Technology, NO-7491 Trondheim, Norway}}


\begin{abstract}
Topological order is important in many aspects of condensed matter physics, and has been extended to bosonic systems. In this Letter we report on the nontrivial topology of the magnon bands in two distinct quantum skyrmion crystals appearing in zero external magnetic field. 
This is revealed by nonzero Chern numbers for some of the bands. 
As a bosonic analog of the quantum anomalous Hall effect, we show that topological magnons can appear in skyrmion crystals without explicitly breaking time-reversal symmetry with an external magnetic field.
By tuning the value of the easy-axis anisotropy at zero temperature, we find eight quantum topological phase transitions signaled by discontinuous jumps in certain Chern numbers. We connect these quantum topological phase transitions to gaps closing and reopening between magnon bands. 

\end{abstract}

\maketitle

\paragraph*{Introduction.}
Topological order in fermionic condensed matter systems lies at the heart of the understanding of the quantum Hall effect (QHE) \cite{TKNN, ChernFermionPRL}, the quantum anomalous Hall effect (QAHE) \cite{Haldane, QAHErev}, the quantum spin Hall effect (QSHE) \cite{KaneMeleQSHE}, and topological insulators (TIs) \cite{HasanKaneRevTI}. The QHE involves explicit time-reversal symmetry breaking by an external magnetic field, while the QSHE and TIs are found in time-reversal-symmetric systems \cite{KaneMeleQSHE, BHZ, MooreBalentsTI, HasanKaneRevTI}. The QAHE is a special case, taking place in systems where time-reversal symmetry is spontaneously, and not explicitly, broken. It is thus a manifestation of the QHE without the need for an external magnetic field \cite{QAHErev}. It was later shown that also bosonic excitations may feature topological properties \cite{ChernBoson1, ChernBoson2, ChernBosonTMI, TopoMagnonRev}. The collective fluctuations of quantum spins, i.e., magnons, have been shown to be topologically nontrivial in magnonic crystals \cite{ChernBoson1}, dipolar magnetic thin films \cite{ChernBoson2}, and in ferromagnets on the honeycomb lattice \cite{RealizeHaldaneKaneMele, ArneTMI, EvenTopoMagnon, owerre2016first, TopoMagnonHoneyFM}. In the model used in Refs.~\cite{EvenTopoMagnon, ArneTMI, RealizeHaldaneKaneMele}, next-nearest neighbor Dzyaloshinskii-Moriya interaction (DMI) realizes a bosonic analog of the Haldane model \cite{Haldane} in a system of insulating spins. 

Even though analogies were proposed \cite{MagnonQHE, MagnonQSH, RealizeHaldaneKaneMele, MagnonChernInsu, TopoMagnonHoneyFM}, topological magnon systems do not show direct equivalences of the QHE, QAHE, and QSHE \cite{EvenTopoMagnon, KaneMeleQSHE, HasanKaneRevTI}. Since bosons obey Bose-Einstein rather than Fermi-Dirac statistics, the Hall conductivity is not quantized \cite{EvenTopoMagnon}. 
The authors of Ref.~\cite{ChernBosonTMI} introduced the bosonic analog of a TI, a topological magnon insulator. The nontrivial topology of the magnon bands gives rise to chiral edge states within a bulk magnon gap. However, since bosonic systems lack the concept of a Fermi surface, the bulk is not guaranteed to be insulating with respect to spin currents \cite{ArneTMI}. 
Still, the chiral edge states resulting from topologically nontrivial magnons allow the creation of magnon currents that are insensitive to backscattering and disorder \cite{ChernBoson1,ChernBoson2}. These hold promising applications such as spin-current splitters, waveguides, and interferometers \cite{ChernBoson1, TopoMagnonRev}. In addition, magnetoelastic coupling leads to chiral phonon transport induced by the topological magnons \cite{EvenTopoMagnon}.

The nontrivial real-space magnetic texture of skyrmions means they are topologically protected \cite{je2020topostab, nagaosaRev}. Therefore, skyrmions received a great deal of interest, and are being explored for applications in magnetic memory technology, unconventional computing, and numerous other applications \cite{nagaosaRev, KlauiRev2016, JonietzCurrentControlSK,skyrmionroadmap,  RacetrackTomasello, RactrackFert, DMIguideSmallJ, ControlSmallJ,  ExchangeFreeJ0, smallSkRW, hsuWriteDeleteSk, yu2017SkRead, skyrmionroadmap, zhangSkAND, SkQubits}. The reciprocal space topology held by the magnon bands in skyrmion crystals (SkXs) has also been explored \cite{RoldanMolina2016, DiazPRL, DiazPRR, WaiznerPhD, ExpTopoMagnonSk}, and a topological phase transition driven by a magnetic field was found in Ref.~\cite{DiazPRR}. Furthermore, evidence of the nontrivial topology of magnon bands in SkXs was observed in an experiment \cite{ExpTopoMagnonSk}. 

In Ref.~\cite{QSkOP}, we explored the quantum fluctuations of the order parameter for quantum SkXs. Quantum skyrmions are skyrmions with such a small size that the continuum limit breaks down, and the quantum nature of the individual spins is not negligible \cite{Sotnikov, Lohani, QSkOP}. 
In this Letter we reveal eight quantum topological phase transitions (QTPTs) \cite{BHZ, QTPT, QTPT_Hamma, QTPT2022} driven by a tunable easy-axis anisotropy in the same quantum SkXs that are explored in Ref.~\cite{QSkOP}. These QTPTs are signaled by discontinuous jumps in the Chern numbers \cite{ChernBoson1} of the magnon bands. Here, we consider QTPTs to be topological phase transitions occurring at zero temperature by tuning a parameter in the Hamiltonian. 
The SkXs we consider are inspired by the observation of a SkX containing nanometer-sized skyrmions in a magnetic monolayer \cite{HeinzeSkX}. 
Since the SkXs are stabilized in zero external magnetic field \cite{QSkOP, HeinzeSkX}, the QTPTs occur in a time-reversal-symmetric model. Instead, the magnetic order of the SkX ground state (GS) spontaneously breaks time-reversal symmetry, allowing nonzero Chern numbers \cite{TopoMagnonRev}. 
In that sense, our skyrmion system is analogous to the QAHE in fermionic systems. This is in contrast to previous studies of topological magnons in SkXs, where time-reversal symmetry is explicitly broken by external magnetic fields \cite{RoldanMolina2016, DiazPRL, DiazPRR, WaiznerPhD, ExpTopoMagnonSk}.

As pointed out in Ref.~\cite{DiazPRL}, the bulk-edge correspondence in not guaranteed unless the finite geometry contains an integer number of unit cells. However, by letting the GS adapt to a strip geometry, the authors of Refs.~\cite{RoldanMolina2016, DiazPRL, DiazPRR} found the expected number and chirality \cite{ChernBoson1} of edge states based on the bulk Chern numbers in SkXs. 
Therefore, we will not explicitly prove the existence of chiral edge states here.
Assuming the validity of the bulk-edge correspondence in a finite geometry, the QTPTs could be used to switch chiral edge states on and off. 

\paragraph*{Model.} \label{sec:model}

\begin{figure*}
    \centering
    \includegraphics[width=0.95\linewidth]{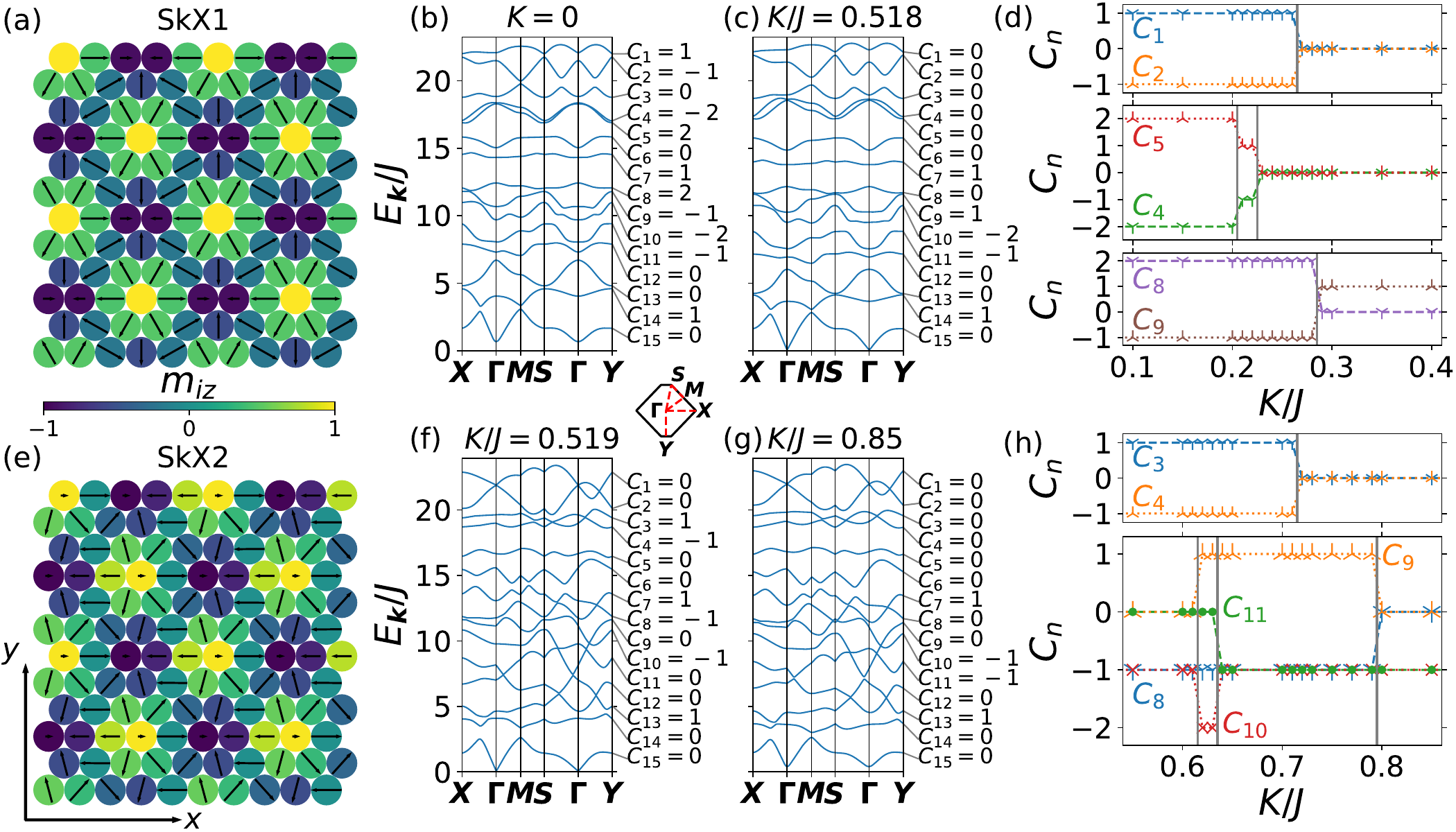}
    \caption{(a) The classical ground state of the skyrmion crystal SkX1, (b) its magnon spectrum at low $K$ along with the Chern numbers of each band, and (c) its magnon spectrum at high $K$ along with the Chern numbers of each band. In (a), colors indicate the $z$ component of the unit vector determining the direction of the spin, $m_{iz}$, while arrows show its projection on the $xy$ plane. The spectra are plotted along the path in the first Brillouin zone (1BZ) that is sketched in the middle. (d) Shows the four quantum topological phase transitions (QTPTs) found in SkX1, where some Chern numbers show discontinuous jumps at the approximate values of $K/J$ shown in gray. The calculated Chern numbers are shown with markers, while dotted or dashed lines are included for illustration. (e)-(h) The same as (a)-(d) but for the distinct skyrmion crystal SkX2. The parameters are $D/J = 2.16, U/J = 0.35$, $S = 1$, (a) $K/J = 0.518$, and (e) $K/J = 0.519$.}
    \label{fig:stateSpectraChern}
\end{figure*}

As in Ref.~\cite{QSkOP}, we use the time-reversal-symmetric Hamiltonian
\begin{equation}
\label{eq:H}
    H = H_{\text{ex}} + H_{\text{DM}} + H_{\text{A}} + H_{4},
\end{equation}
where
\begin{equation}
    H_{\text{ex}}  = -J\sum_{\langle ij \rangle} \boldsymbol{S}_{i}\cdot \boldsymbol{S}_{j},
\end{equation}
\begin{equation}
    H_{\text{DM}} = \sum_{\langle ij \rangle} \boldsymbol{D}_{ij} \cdot (\boldsymbol{S}_i \cross \boldsymbol{S}_j),
\end{equation}
\begin{equation}
        H_{\text{A}} = - K\sum_i S_{iz}^2, 
\end{equation}
\begin{align}
    H_4 =& U\sum_{ijkl}^\diamond \big[(\boldsymbol{S}_i \cdot \boldsymbol{S}_j)(\boldsymbol{S}_k \cdot \boldsymbol{S}_l) + (\boldsymbol{S}_i \cdot \boldsymbol{S}_l)(\boldsymbol{S}_j \cdot \boldsymbol{S}_k) \nonumber \\
    &\mbox{\qquad\quad}-(\boldsymbol{S}_i \cdot \boldsymbol{S}_k)(\boldsymbol{S}_j \cdot \boldsymbol{S}_l)\big].
\end{align}
The spin operator $\boldsymbol{S}_i$, with magnitude $S$, pertains to lattice site $i$ on the triangular lattice. We consider a nearest-neighbor ferromagnetic exchange interaction, $J>0$. In Ref.~\cite{HeinzeSkX}, DMI between Fe atoms on the surface originates with the strong spin-orbit coupling from the Ir atoms \cite{KlauiRev2016}. We assume a similar effect in our model and set the DMI vector to $\boldsymbol{D}_{ij} = D \hat{r}_{ij} \cross \hat{z}$, where $\hat{r}_{ij}$ is a unit vector from site $i$ to site $j$. Like in Ref.~\cite{QSkOP}, we will discuss a tunable easy-axis anisotropy, $K$, motivated by the findings of Refs.~\cite{TuneK_PRB, TuneK} where it was shown that applying mechanical strain can tune the magnetic anisotropy. The Hamiltonian also contains the four-spin interaction $H_4$, acting between four sites that are oriented counterclockwise and make diamonds of minimal area \cite{HeinzeSkX, H4PRB}. The reduced Planck's constant $\hbar$ and the lattice constant $a$ are set to $1$.

We refer to Refs.~\cite{QSkOP, Suppl} for details of the two distinct SkX GSs, which are separated by a quantum phase transition (QPT) at $K = K_t$ with $K_t/J \in (0.518, 0.519)$. The classical GSs of SkX1 and SkX2 are shown in Fig.~\ref{fig:stateSpectraChern}(a,e) for $K/J = 0.518$ and $K/J = 0.519$, respectively. 
By including quantum corrections in a calculation of the expectation value of the Hamiltonian, it is found that $\langle H \rangle$ is lower than the classical GS energy. Hence, quantum skyrmions are energetically preferred over their corresponding classical GSs. \cite{QSkOP, Suppl}. 
The introduction of the Holstein-Primakoff transformation via rotated coordinates \cite{HProtation_2009} involves approximations whose validity are discussed in Refs.~\cite{QSkOP, Suppl}. Possible corrections to our predictions due to the ignored magnon-magnon interactions are discussed in Refs.~\cite{DiazPRL, TopoMagnonRev}. 
Dipolar interactions were shown to affect the high-energy magnon bands of SkXs in Ref.~\cite{WaiznerPhD}. 
Since we consider a magnetic monolayer, dipolar interactions are not expected to have significant effects \cite{KlauiRev2016, HeinzeSkX}. The study of dipolar interactions is also beyond the scope of this Letter. For these reasons, they are excluded from our model.

The diagonalization of the system to obtain the magnon bands \cite{COLPA} is shown in detail in Refs.~\cite{QSkOP, Suppl}. We take the transformation matrices $T_{\boldsymbol{k}}$ and the magnon bands $E_{\boldsymbol{k},n}$ as inputs in this Letter. The 15 magnon bands are numbered from top to bottom in terms of energy. The first Brillouin zone (1BZ) is the same for all 15 sublattices in both SkX1 and SkX2 and we define the points $\boldsymbol{\Gamma} = (0,0)$, $\boldsymbol{X} = (52\pi/135, 0)$, $\boldsymbol{M} = (\pi/5, \pi/3\sqrt{3})$, $\boldsymbol{S} = (2\pi/135, 2\pi/3\sqrt{3})$, and $\boldsymbol{Y} = (0, -2\pi/3\sqrt{3})$ in the 1BZ \cite{QSkOP, Suppl}. These points and the 1BZ are sketched in Fig.~\ref{fig:stateSpectraChern}.

\paragraph*{Chern numbers.} \label{sec:Chern}
Let $\Gamma_n$ be a matrix whose $n$th diagonal element is $1$ and all other matrix elements are zero, $\Gamma_{n,i,j} = \delta_{in}\delta_{jn} $. From this, we define a projection matrix $P_{\boldsymbol{k},n} = T_{\boldsymbol{k}}^{-1} \Gamma_n T_{\boldsymbol{k}}$. The bosonic nature of the magnons is encoded in the paraunitary transformation matrix \cite{COLPA, QSkOP, Suppl}. Then the Berry curvature of the $n$th band is given by \cite{ChernBoson1, ChernFermionPRL}
\begin{equation}
\label{eq:berry}
    B_n(\boldsymbol{k}) = i\epsilon_{\mu\nu} \Tr[(\delta_{k_\mu}P_{\boldsymbol{k},n})P_{\boldsymbol{k},n}(\delta_{k_\nu} P_{\boldsymbol{k},n})],
\end{equation}
where $\epsilon_{\mu\nu}$ is the Levi-Civita tensor and $\mu,\nu \in \{x,y\}$. The Chern number of the $n$th band is its Berry curvature integrated over the 1BZ
\begin{equation}
\label{eq:chern}
    C_n = \frac{1}{2\pi}\int_{\text{1BZ}} d \boldsymbol{k} B_n(\boldsymbol{k}).
\end{equation}
It can be shown that the Chern numbers are integers, given that the bands are isolated \cite{ChernBoson1}. Here, we calculate the Chern numbers using numerical approximations to the integral. We consider equally spaced discretizations and adaptive quadratures \cite{Suppl, AdaptQuad}. 
When the numerical results are found to approach integers upon increasing the density of $\boldsymbol{k}$ points, we present the Chern numbers as integers.

\paragraph*{Quantum topological phase transitions.}
In Fig.~\ref{fig:stateSpectraChern}(b,c) we show the magnon spectrum in SkX1 for $K=0$ and for $K/J = 0.518$, i.e., close to the QPT to SkX2. The Chern numbers of the 15 bands are given at both values of $K$ and it is clear that some of them have changed due to the change in easy-axis anisotropy. In Fig.~\ref{fig:stateSpectraChern}(d) we plot these as a function of $K$, revealing four QTPTs. $E_{\boldsymbol{k}, 4}$ and $E_{\boldsymbol{k}, 5}$ first cross at the $\boldsymbol{Y}$ point for the specific value $K = K_1$, where $K_1/J$ is in the interval $(0.20, 0.21)$. They also cross at the $\boldsymbol{\Gamma}$ point for $K/J = K_2/J \in (0.22, 0.23)$. The gap between the bands closes and reopens, and their Chern numbers change, signaling QTPTs. $E_{\boldsymbol{k}, 1}$ and $E_{\boldsymbol{k}, 2}$ cross at the $\boldsymbol{Y}$ point for $K/J = K_3/J \in (0.26, 0.27)$. The two Chern numbers annihilate, and both bands are topologically trivial for $K>K_3$. Finally, the gap between $E_{\boldsymbol{k}, 8}$ and $E_{\boldsymbol{k}, 9}$ closes between the $\boldsymbol{\Gamma}$ point and the $\boldsymbol{X}$ point for $K/J = K_4/J \in (0.28, 0.29)$. Only $E_{\boldsymbol{k}, 9}$ remains topologically nontrivial when the gap reopens for $K>K_4$. 
In SkX1, the magnon band with lowest energy is topologically trivial, while the band with second lowest energy is topologically nontrivial. For ferromagnetic SkXs in an external magnetic field, the band with third lowest energy is topologically nontrivial while the two bands with lower energy are topologically trivial \cite{RoldanMolina2016, DiazPRR, WaiznerPhD, ExpTopoMagnonSk}.

Fig.~\ref{fig:stateSpectraChern}(e,f) shows the magnon spectrum in SkX2 for $K/J=0.519$ and $K/J = 0.85$. Despite the plethora of closely avoided crossings, all the bands are isolated at these values of $K$, and all 15 Chern numbers are well defined. Notice that in all cases the sum of the Chern numbers of all bands is zero, as expected \cite{ChernBoson1}. It is clear that the Chern numbers have changed from the spectrum of SkX1 at $K/J = 0.518$ to the spectrum of SkX2 at $K/J = 0.519$. 
We do not view this as a QTPT, since it is not due to gaps closing and reopening in the magnon spectrum. Rather, the magnon spectra are different from the outset, since they arise from two distinct SkXs.

In Fig.~\ref{fig:stateSpectraChern}(g) we plot the Chern numbers that change when tuning $K$ in SkX2. We find four QTPTs. The gap between $E_{\boldsymbol{k},9}$ and $E_{\boldsymbol{k},10}$ closes between $\boldsymbol{\Gamma}$ and $\boldsymbol{Y}$ for $K/J = K_5/J \in (0.61, 0.62)$. Once the gap reopens, $E_{\boldsymbol{k},9}$ has become topologically nontrivial, while $C_{10} = -1$ has jumped to $C_{10} = -2$. $E_{\boldsymbol{k},10}$ and $E_{\boldsymbol{k},11}$ cross between $\boldsymbol{\Gamma}$ and $\boldsymbol{Y}$ for $K/J = K_6/J \in (0.63,0.64)$. $C_{10}$ jumps back to $-1$, allowing $E_{\boldsymbol{k},11}$ to become topologically nontrivial for $K>K_6$. The gap between $E_{\boldsymbol{k},3}$ and $E_{\boldsymbol{k},4}$ closes between $\boldsymbol{\Gamma}$ and $\boldsymbol{Y}$ for $K/J = K_7/J \in (0.71, 0.72)$. Once the gap reopens at $K>K_7$ they are both topologically trivial. Finally, $E_{\boldsymbol{k},8}$ and $E_{\boldsymbol{k},9}$ cross at $\boldsymbol{k} \approx (\pm 0.32, 0.10)$ for $K/J = K_8/J \in (0.79, 0.80)$ and are left topologically trivial for $K > K_8$.

\begin{figure*}
    \centering
    \includegraphics[width=0.95\linewidth]{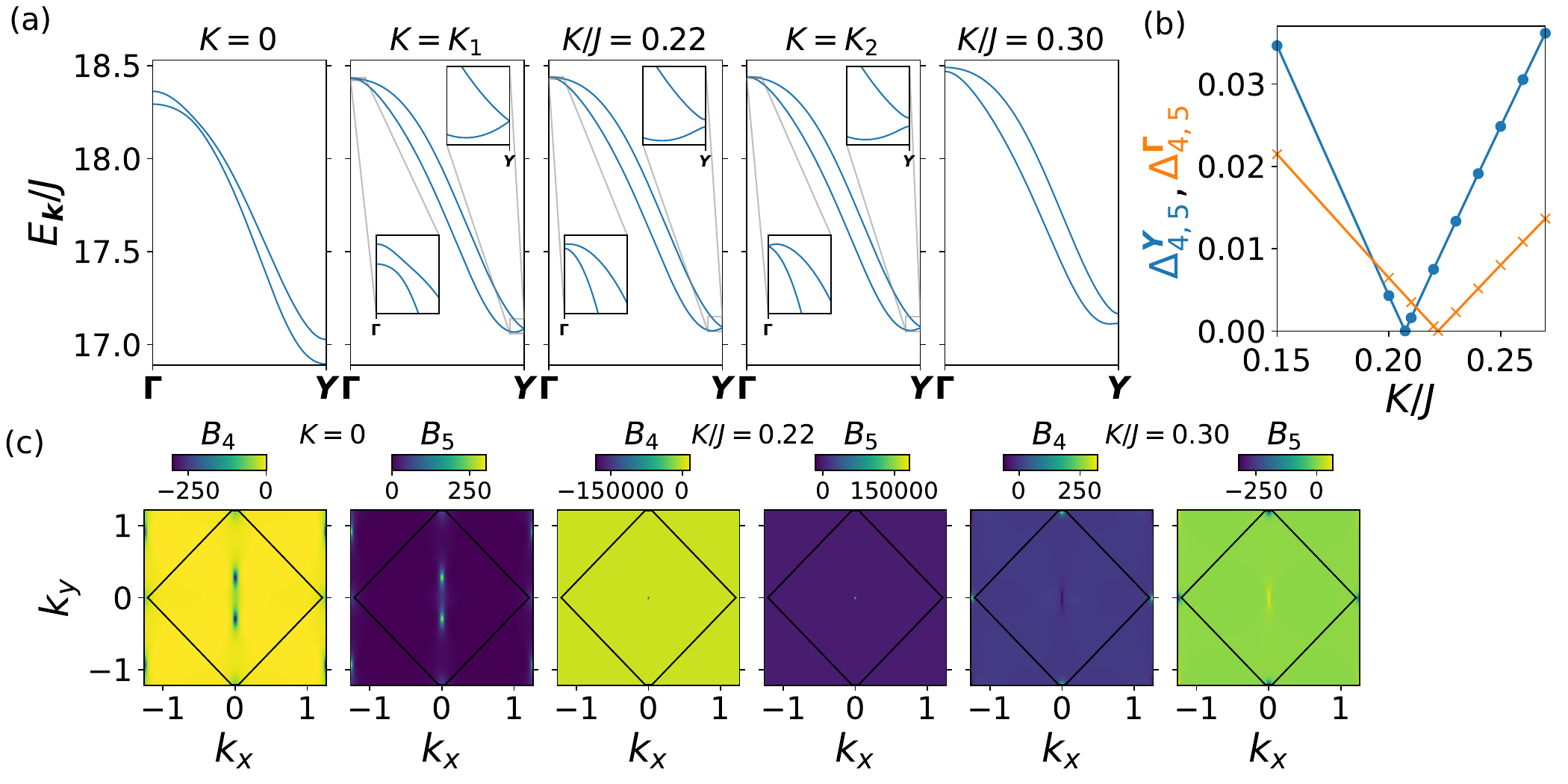}
    \caption{(a) Plots of $E_{\boldsymbol{k}, 4}$ and $E_{\boldsymbol{k}, 5}$ in SkX1 at varying $K$ showing how the gap at $\boldsymbol{Y}$ closes at $K=K_1$ where there is a QTPT, and $C_4 = -2, C_5 = 2$ at $K< K_1$ goes to $C_4 = -1, C_5 = 1$ at $K_1 < K < K_2$ once the gap reopens. Also, the gap closes at the $\boldsymbol{\Gamma}$ point for $K = K_2$, and once it reopens, the two bands are topologically trivial. (b) The gap between $E_{\boldsymbol{k}, 4}$ and $E_{\boldsymbol{k}, 5}$ at the $\boldsymbol{Y}$ ($\boldsymbol{\Gamma}$) point is shown in blue (orange), with circles (crosses) at the calculated values. (c) The Berry curvatures of bands $E_{\boldsymbol{k}, 4}$ and $E_{\boldsymbol{k}, 5}$ shown for the three values of $K$ in (a) where the bands are isolated. The 1BZ is indicated in black and the Berry curvatures are plotted with 201 points in each direction. The parameters are $D/J = 2.16, U/J = 0.35$, and $S = 1$.}
    \label{fig:gapcloseBerry}
\end{figure*}

\paragraph*{Gap closing and Berry curvature.} \label{sec:Berry}

In Fig.~\ref{fig:gapcloseBerry}(a) we go into detail of the two band crossings of $E_{\boldsymbol{k}, 4}$ and $E_{\boldsymbol{k}, 5}$ in SkX1. For $K<K_1$ we have $C_4 = -2, C_5 = 2$. Then, at $K=K_1$ the gap between $E_{\boldsymbol{k}, 4}$ and $E_{\boldsymbol{k}, 5}$ at the $\boldsymbol{Y}$ point closes, $\Delta_{4,5}^{\boldsymbol{Y}} = 0$, and the two Chern numbers are undefined \cite{ChernBoson1}. For $K>K_1$ the gap reopens and $C_4 = -1, C_5 = 1$, i.e., the bands remain topologically nontrivial. For $K=K_2$ the gap closes at the $\boldsymbol{\Gamma}$ point, $\Delta_{4,5}^{\boldsymbol{\Gamma}} = 0$. When the gap reopens for $K>K_2$ both bands have become topologically trivial. The dependence of these two gaps on the easy-axis anisotropy is shown in Fig.~\ref{fig:gapcloseBerry}(b). It appears the two gaps close with an approximately linear dependence on $K$.

The Berry curvatures of $E_{\boldsymbol{k}, 4}$ and $E_{\boldsymbol{k}, 5}$ are shown in Fig.~\ref{fig:gapcloseBerry}(c). At $K =0$, $B_4(\boldsymbol{k})$ has extended negative valleys, giving rise to a negative Chern number. For $K/J = 0.22$, $\Delta_{4,5}^{\boldsymbol{\Gamma}}$ is small and so there is a sharp negative valley in $B_4(\boldsymbol{k})$ around $\boldsymbol{k} = \boldsymbol{\Gamma}$. Again, this gives rise to a negative Chern number. At $K/J = 0.3$, the Berry curvature contains both positive peaks and negative valleys, which cancel each other out in the integral and lead to zero Chern number. The arguments are similar for $B_5(\boldsymbol{k})$ and $C_5$. From these figures, it is clear that the gap closings involve an exchange of Berry curvature between the bands. Also, the Berry curvature of a given band has its largest absolute values where the band has the smallest gap to neighboring bands. Similar figures and arguments can be extended to the remaining six QTPTs discussed in this Letter. Notice that each time two bands cross and undergo a QTPT, the sum of their Chern numbers is preserved, as expected \cite{ChernFermionPRL}.

\paragraph*{Predicted edge states.}
The predicted number of edge states within the gap between the bands $E_{\boldsymbol{k},n}$ and $E_{\boldsymbol{k},n+1}$ is 
\begin{equation}
    \nu_n = \sum_{n' = n+1}^{15} C_{n'}. 
\end{equation}
The chiral edge states propagate clockwise (counterclockwise) for positive (negative) $\nu_n$ \cite{ChernBoson1}. For instance, we predict a clockwise edge state within the bulk band gap between $E_{\boldsymbol{k},13}$ and $E_{\boldsymbol{k},14}$ in SkX1. 
Let SkX3 be the result of applying the time-reversal operator, i.e., flipping all spins, to SkX1. In Ref.~\cite{QSkOP}, we mentioned that since the Hamiltonian in Eq.~\eqref{eq:H} is time-reversal symmetric, SkX1 and SkX3 are degenerate in energy. It was also mentioned that the two states can appear concurrently, separated by domain walls. SkX3 has the same magnon spectrum as SkX1, while the Chern numbers change sign. At the interface between two topologically nontrivial systems $A$ and $B$, with $\nu_n = \nu_A$ and $\nu_{n} = \nu_B$ in the same energy interval, one expects $|\nu_A-\nu_B|$ edge states \cite{RoldanMolina2016}. Therefore, along a domain wall between SkX1 and SkX3, we predict two chiral edge states within the gap between $E_{\boldsymbol{k},13}$ and $E_{\boldsymbol{k},14}$.
\paragraph*{Conclusion.} \label{sec:Con}
We found eight quantum topological phase transitions in two distinct skyrmion crystals that are stabilized in a time-reversal-symmetric model. Time-reversal symmetry is spontaneously broken by the magnetic ordering of the skyrmions, and therefore nonzero Chern numbers of the magnon bands are possible. This is a bosonic analog of the quantum anomalous Hall effect.
The quantum topological phase transitions, driven by a tunable easy-axis anisotropy at zero temperature, are signaled by jumps in the Chern numbers. We illustrated how the closing and subsequent reopening of the gaps between magnon bands leads to these jumps in the Chern numbers, and how the Berry curvature depends on these gaps. 

\paragraph*{Acknowledgments.}
We acknowledge funding from the Research Council of Norway through its Centres of Excellence funding scheme, Project No.~262633, ``QuSpin," and through Project No.~323766, ``Equilibrium and out-of-equilibrium quantum phenomena in superconducting hybrids with antiferromagnets and topological insulators."

\bibliography{main.bbl}

\clearpage
\onecolumngrid
\allowdisplaybreaks

\renewcommand{\thefigure}{S\arabic{figure}}
\renewcommand{\theHfigure}{S\arabic{figure}}
\setcounter{figure}{0}  


\renewcommand{\thetable}{S\Roman{table}}
\setcounter{table}{0}  

\renewcommand{\theequation}{S\arabic{equation}}
\setcounter{equation}{0}  

\renewcommand{\thesection}{S\arabic{section}}
\renewcommand{\thesubsection}{\thesection.\arabic{subsection}}
\renewcommand{\thesubsubsection}{\thesubsection.\arabic{subsubsection}}
\makeatletter
\renewcommand{\p@subsection}{}
\renewcommand{\p@subsubsection}{}
\makeatother

\phantomsection
\begin{large}
\begin{center}
    \textbf{Supplemental material for ``Quantum topological phase transitions in skyrmion crystals''} \label{sec:Suppl}
\end{center}
\end{large}



\tableofcontents

\section{Introduction}
Sections \ref{sec:GS} and \ref{sec:HProtation} of this supplemental material repeat relevant details in Ref.~\cite{QSkOP} that are left out of the main text of ``Quantum topological phase transitions in skyrmion crystals''. Section \ref{sec:GS} concerns details of obtaining the classical ground state numerically, while Sec.~\ref{sec:HProtation} details the Holstein-Primakoff (HP) approach toward finding the magnon energy spectrum. In both sections, more details are included than what was previously presented in Ref.~\cite{QSkOP}. Section \ref{sec:Cherndetail} gives details of the numerical calculation of Chern numbers.

\section{Classical ground state} \label{sec:GS}

\subsection{Periodicity}
The classical Hamiltonian is obtained from the Hamiltonian in Eq.~(1) of the main text by setting $\boldsymbol{S}_i = S \boldsymbol{m}_i$. Here, $\boldsymbol{m}_i$ is a unit vector along the direction of the spin at lattice site $i$, while $S$ is the uniform spin magnitude. This gives
\begin{equation}
\label{eq:Hclass}
    H(\{\boldsymbol{m}_i\}) = H_{\text{ex}} + H_{\text{DM}} + H_{\text{A}} + H_{4},
\end{equation}
where
\begin{equation}
    H_{\text{ex}}  = -JS^2 \sum_{\langle ij \rangle} \boldsymbol{m}_{i}\cdot \boldsymbol{m}_{j},
\end{equation}
\begin{equation}
    H_{\text{DM}} = S^2 \sum_{\langle ij \rangle} \boldsymbol{D}_{ij} \cdot (\boldsymbol{m}_i \cross \boldsymbol{m}_j),
\end{equation}
\begin{equation}
    H_{\text{A}} = - KS^2 \sum_i m_{iz}^2, 
\end{equation}
\begin{align}
    H_4 =& US^4 \sum_{ijkl}^\diamond \big[(\boldsymbol{m}_i \cdot \boldsymbol{m}_j)(\boldsymbol{m}_k \cdot \boldsymbol{m}_l) + (\boldsymbol{m}_i \cdot \boldsymbol{m}_l)(\boldsymbol{m}_j \cdot \boldsymbol{m}_k) \nonumber \\
    &\mbox{\qquad\quad}-(\boldsymbol{m}_i \cdot \boldsymbol{m}_k)(\boldsymbol{m}_j \cdot \boldsymbol{m}_l)\big].
\end{align}
Minimizing $H(\{\boldsymbol{m}_i\})$ with respect to $\{\boldsymbol{m}_i\}$ will reveal the classical ground state (GS) of the system. Potential candidates are collinear states such as ferromagnetic states, coplanar states such as helical states, or noncoplanar states such as skyrmions. With $U \neq 0$ it turns out ferromagnetic and helical states will be disfavored, making skyrmion crystal (SkX) GSs more likely. The Hamiltonian is far too complicated to obtain an analytic solution for the GS, and so we must resort to numerical methods. Additionally, we wish to study a bulk system with a periodic magnetic state, in order to introduce the HP approach \cite{HProtation_2009}. To obtain results within a reasonable amount of time, a limited number of spins can be included in the simulations. If the periodicity of the ground state is unknown, any chosen lattice size with periodic boundary conditions (PBCs) will introduce finite-size effects. Hence, it is a major advantage to know the periodicity of the GS, before searching for the optimal state with that periodicity.

The periodicity of the GS will depend on the parameters in the model. Our approach is to tune the parameters to ensure that a SkX with the same periodicity as the best commensurate approximation to the one observed experimentally in Ref.~\cite{HeinzeSkX}, is the GS of the system. From the supplementary information of Ref.~\cite{HeinzeSkX} we take the skyrmion constructor $\boldsymbol{m}_i = (\sin\Tilde{\phi}_i \cos\Tilde{\theta}_i/|\cos\Tilde{\theta}_i|, \cos\Tilde{\phi}_i \sin\Tilde{\theta}_i, \cos\Tilde{\phi}_i \cos\Tilde{\theta}_i)$, where $\Tilde{\phi}_i = \boldsymbol{Q}_M \cdot \boldsymbol{r}_i, \Tilde{\theta}_i = \boldsymbol{Q}_K \cdot \boldsymbol{r}_i, \boldsymbol{Q}_M =  2\pi \hat{x}/\lambda_x, \boldsymbol{Q}_K =  2\pi \hat{y}/ (\lambda_y\sqrt{3}/2)$. $\lambda_x$ is the periodicity in the $x$ direction in terms of lattice sites, while $\lambda_y$ is the periodicity in the $y$ direction in terms of lattice chains. We name these trial states SkXt. For any rational numbers $\lambda_x$ and $\lambda_y$ it is possible to construct a finite-sized lattice with correct PBCs in order to calculate the energy per site.

\begin{figure}
    \centering
    \includegraphics[width=0.99\linewidth]{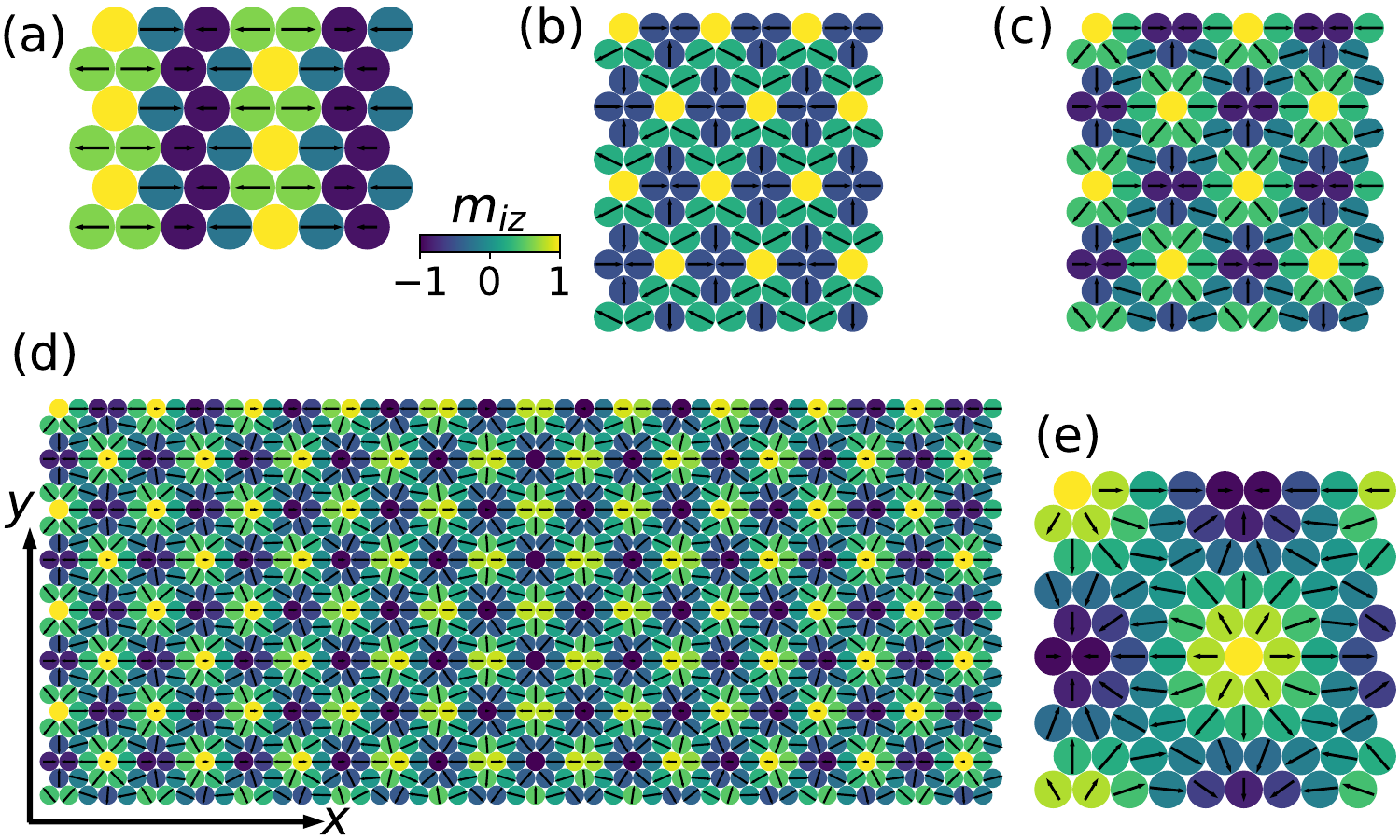}
    \caption{SkXt for varying periodicities. Colors indicate $m_{iz}$, while the projection of $\boldsymbol{m}_i$ on the $xy$ plane is shown with arrows. (a) SkXt with $\lambda_x = 3.5, \lambda_y = 1$ is a helical state. SkXt with (b) $\lambda_x = 3, \lambda_y = 6$, (c) $\lambda_x = 5, \lambda_y = 6$, (d) $\lambda_x = 4.9, \lambda_y = 6$, and (e) $\lambda_x = 9, \lambda_y = 10$ are skyrmion states. Their energies are given in Table \ref{tab:periodicityenergy}. }
    \label{fig:supplstates}
\end{figure}
\begin{table}[ht]
    \centering
    \caption{The four individual contributions to the Hamiltonian and the total Hamiltonian for SkXt states given per lattice site. Some of the states are shown in Fig.~\ref{fig:supplstates}, and a reference to the corresponding subfigure is given in the column titled ``Figure''. For each choice of $\lambda_x, \lambda_y$ we set up a lattice with a size such that periodic boundary conditions can be implemented like a torus. One such choice is shown for each of the states that are plotted in Fig.~\ref{fig:supplstates}. The parameters are $D/J = 2.16$, $U/J = 0.35$, $K/J = 0.10$, and $S=1$. }
    \begin{tabular}{D{.}{.}{1}D{.}{.}{1}D{.}{.}{3}D{.}{.}{3}D{.}{.}{3}D{.}{.}{3}D{.}{.}{3}cc}
        \hline\hline
        \multicolumn{1}{c}{$\lambda_x$} & \multicolumn{1}{c}{$\lambda_y$} & \multicolumn{1}{c}{$H_{\text{ex}}/NJ$} & \multicolumn{1}{c}{$H_{\text{DM}}/NJ$} & \multicolumn{1}{c}{$H_{\text{A}}/NJ$} & \multicolumn{1}{c}{$H_4/NJ$} & \multicolumn{1}{c}{$H_{\text{SkXt}}/NJ$} & Figure & Description \\
        \hline
        1.0  &  2.0 & -6.000 &  0.000 & -0.100 &  4.200 & -1.900 &                          & Ferromagnet  \\
        3.5  &  1.0 & -2.049 & -7.589 & -0.050 &  4.200 & -5.488 & \ref{fig:supplstates}(a) & Helical  \\
        3.0  &  6.0 &  0.167 & -5.361 & -0.025 &  0.875 & -4.345 & \ref{fig:supplstates}(b) & SkX  \\
        4.9  &  6.0 & -1.905 & -6.219 & -0.025 &  0.892 & -7.256 & \ref{fig:supplstates}(d) & SkX  \\
        5.0  &  6.0 & -1.966 & -6.207 & -0.025 &  0.939 & -7.259 & \ref{fig:supplstates}(c) & SkX  \\
        5.1  &  6.0 & -2.025 & -6.193 & -0.025 &  0.986 & -7.256 &                          & SkX  \\
        5.0  &  6.1 & -2.008 & -6.043 & -0.025 &  0.983 & -7.093 &                          & SkX  \\
        9.0  & 10.0 & -4.180 & -4.637 & -0.025 &  3.029 & -5.814 & \ref{fig:supplstates}(e) & SkX  \\
        \hline\hline
    \end{tabular}
    \label{tab:periodicityenergy}
\end{table}


We show a selection of SkXt states in Fig.~\ref{fig:supplstates}. In Table \ref{tab:periodicityenergy} we show the total energy of SkXt as well as the four individual contributions for various choices of peridicity.
The exchange coupling, $H_{\text{ex}}$, prefers a ferromagnetic GS. Among the helical and skyrmion-like states $H_{\text{ex}}$ prefers large periodicities since then neighboring spins are more aligned. The Dzyaloshinskii-Moriya interaction (DMI), $H_{\text{DM}}$, favors helical states, especially the helical state with $\lambda_x \approx 3.4, \lambda_y = 1$.
The helical state with $\lambda_x = 3.5, \lambda_y = 1$ is shown in Fig.~\ref{fig:supplstates}(a). 
The easy-axis anisotropy, $H_{\text{A}}$, has no effect on the periodicity unless it becomes the dominant energy in the system. It prefers collinear states along the $z$ axis. Coplanar states are also preferred over noncoplanar states. The four-spin interaction, $H_4$, disfavors collinear and coplanar states. It prefers noncoplanar SkXt states with small periodicities. Along with the DMI term, $H_4$ is instrumental in stabilizing SkXs with small peridicities in our model. With parameters tuned to $D/J = 2.16$ and $US^2/J = 0.35$, 
the total Hamiltonian is minimized by the SkXt state with $\lambda_x = 5, \lambda_y = 6$ which is shown in Fig.~\ref{fig:supplstates}(c). We checked that changing either periodicity by $\pm 0.001$ yielded states with higher energy. 

A system where the SkXt in Fig.~\ref{fig:supplstates}(b) is the GS may allow studying even smaller skyrmions. Meanwhile, the state in Fig.~\ref{fig:supplstates}(e) shows larger skyrmions. States such as that in Fig.~\ref{fig:supplstates}(d), with noninteger periodicity, may be used to study an incommensurate SkX. An incommensurate SkX was observed in Ref.~\cite{HeinzeSkX}. To limit the number of spins in the magnetic unit cell, and hence the number of magnon bands, we chose to work with a commensurate SkX with $\lambda_x = 5, \lambda_y = 6$ as the preferred periodicity.

\begin{figure}
    \centering
    \includegraphics[width = 0.4\textwidth]{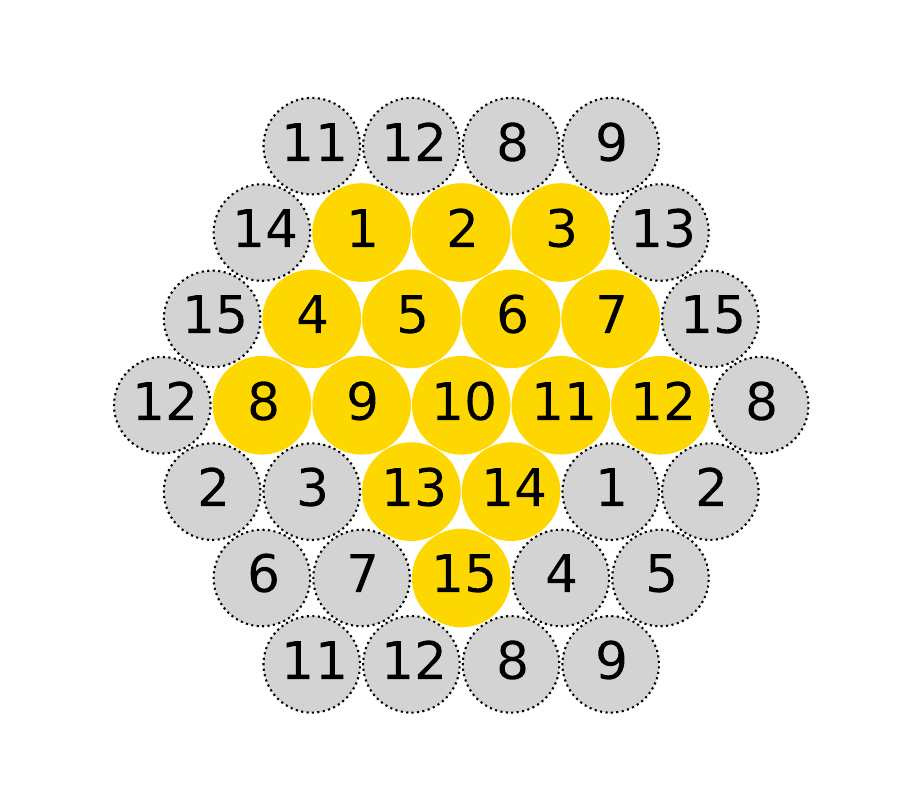}
    \caption{Gold lattice sites show the 15-site cluster making up the magnetic unit cell, and the adopted numbering of the 15 sublattices. The grey lattice sites surrounded by dotted lines are used to indicate the periodic boundary conditions (PBCs), with the number indicating which sublattice the periodic coupling goes to. This form of PBC is like a twisted torus \cite{PBChexagon}. Compare to Fig.~\ref{fig:supplstates}(c) and let, e.g., sublattice 10 be the central spin pointing up.}
    \label{fig:MWC}
\end{figure}

\subsection{Obtaining the ground states}
We now assume that with the parameters tuned to the values above, the GS is a SkX with a periodicity of $\lambda_x = 5, \lambda_y = 6$. We can then perform simulations with lattice sizes $5\cross 6, 10 \cross 12,$ etc., or, we can define a 15 site magnetic unit cell containing the only 15 unique spins, and implement PBCs similar to a twisted torus \cite{PBChexagon}; see Fig.~\ref{fig:MWC}. The latter approach gave the states with lowest energy, while the first choices gave states that visually were periodic as assumed, but where the periodicity was not exact and the resultant energy slightly higher. Hence they serve as nice tests, but were not used to obtain any results. 


We implemented a Monte Carlo simulated annealing approach \cite{simulatedannealing, DiazPRL, DiazPRR} to search for potential GSs. The algorithm is
\begin{enumerate}
    \item Create a random starting distribution $\{\theta_i, \phi_i\}$ specifying $\boldsymbol{m}_i = (\sin\theta_i\cos\phi_i,$ $\sin\theta_i\sin\phi_i,$ $\cos\theta_i)$. The inclination $\theta_i$ and the azimuth $\phi_i$ are the familiar angles in spherical coordinates. 
    \item Pick a random site $j$ and update $\theta_j, \phi_j$. \label{item:update}
    \item Calculate the change in energy, $\Delta H$, and $W = e^{-\beta\Delta H}$, where $\beta$ is the inverse temperature.
    \item Pick a random number, $r$, between $0$ and $1$. If $W > r$ accept the new state.
    \item Decrease temperature, and repeat from step \ref{item:update} for a chosen number of repetitions, usually quite large.
\end{enumerate} 
Several thermalization schemes were attempted in order to gradually cool the system from a high initial temperature to a low temperature compared to the relevant energy scales in the system. The idea is that starting with high temperature should reduce the risk of getting stuck in local minima. Additionally running many repetitions with a very low temperature after the initial thermalization aims to obtain the lowest energy state, i.e., the GS of the system. Alternatively, one can start with a state that is assumed to be similar to the GS of the system. One such choice would be SkXt at $\lambda_x = 5, \lambda_y = 6$. In the latter case, the starting temperature should also be low. If not, the initial spin updates will just bring the system out of the SkXt state into what is essentially a high energy random state.

We also employed a self-consistent iteration approach \cite{dosSantosPRB, RoldanMolina, RoldanMolina2016}, which yielded states with lower energy than any we obtained using Monte Carlo simulated annealing. Therefore, all results are calculated from numerical GSs obtained using self-consistent iteration. The algorithm is \cite{dosSantosPRB, QSkOP}
\begin{enumerate}
    \item Create a random starting distribution $\{\theta_i, \phi_i\}$ specifying $\{\boldsymbol{m}_i \}$. Alternatively, start from a judiciously chosen distribution based on the suspected GS.
    \item Assume $n \in \mathbb{N}$ iterations have been completed. Then, for each spin, calculate the magnetic torques $T_i^\theta = \partial_{\theta_i^n}{H}$ and $T_i^\phi = \partial_{\phi_i^n}{H}$. Use these to set new angles $\theta_i^{n+1} = \theta_i^{n} -\alpha T_i^\theta$ and $\phi_i^{n+1} = \phi_i^{n} -\alpha T_i^\phi ,$ where $\alpha$ is the mixing parameter \cite{dosSantosPRB}. We used $\alpha J = 0.001$ and $0.0001$. \label{item:updateit}
    \item Repeat step \ref{item:updateit} until self-consistency is reached.
\end{enumerate} 

We checked that values of $K/J$ in the interval $[0,1]$ had no effect on the periodicity of the SkX GS. It is likely this interval can be extended, but we believe large magnitudes of $K$ will eventually change the periodicity of the GS. An interesting effect was found at $K = K_t$, where we determined that $K_t/J$ is somewhere in the interval $(0.518, 0.519)$. There, the SkX GS changes nature, but not periodicity. For $0 \leq K/J \leq 0.518$ the lowest energy state is named SkX1, which is shown in Fig.~1(a) of the main text. Tuning $K$ in that region only adjusts the $z$ component of all spins gradually. However, for $0.519 \leq K/J \leq 1$ the center of the skyrmion relocates along the $x$-axis by approximately one quarter lattice constant. States where the center relocates to the left or right are degenerate in energy. We chose to focus on the state SkX2 shown in Fig.~1(e) of the main text, where the center has relocated to the left. The lower symmetry of SkX2 compared to SkX1 distinguishes it as a different phase. The phase transition between SkX1 and SkX2 occurs at zero temperature by tuning a parameter in the Hamiltonian, and is therefore a quantum phase transition (QPT) \cite{QSkOP}. 

SkX1 and SkX2 are found to have lower energy than the trial state SkXt that was used to determine the periodicity. New GSs are found at each considered value of $K/J$. To double check the proposed GSs we also started from random states on larger lattices of varying sizes using Monte Carlo simulated annealing. Due to the complicated energy landscape, Monte Carlo simulations might get stuck in local minima. To check if this prevented finding helical states, we also searched specifically for the best possible coplanar state, without finding any states with lower energy than SkX1 and SkX2. 

\subsubsection{Net magnetization in ground states}
We define the GS net magnetizations as
\begin{equation}
    \overline{m}_\alpha = \frac{1}{N}\sum_i m_{i\alpha}.
\end{equation}
In SkX1 $\overline{m}_x = \overline{m}_y = 0$ within numerical accuracy, while $\overline{m}_z$ is small and negative. The net magnetization is $\overline{m}_z \approx -0.002$ at low $K/J$, with decreasing magnitude for increasing $K/J$. In SkX2 $\overline{m}_y = \overline{m}_z = 0$ within numerical accuracy, while $\overline{m}_x$ is small and negative. The net magnetization is $\overline{m}_x \approx -0.014$ with a minor dependence on $K$. Hence, both states have a small net magnetization even though there is no external magnetic field. Apparently, a small net magnetization allows the formation of a state with lower energy than any state with zero net magnetization. 


\section{Holstein-Primakoff approach} \label{sec:HProtation}

\subsection{Inserting rotated coordinates}
Following Ref.~\cite{HProtation_2009} we introduce a local orthonormal frame $\{\hat{e}_1^i, \hat{e}_2^i, \hat{e}_3^i\}$ with $\hat{e}_3^i = \boldsymbol{m}_i$. I.e., a coordinate system where the third axis points along the magnetization in the classical GS. We let $\hat{e}_\alpha^i = (R_{\alpha\beta}^i)^{-1} \hat{r}_\beta$, where $\hat{r}_\beta$ are the Cartesian axes.
Inverting we get $\hat{r}_\alpha = R_{\alpha\beta}^i \hat{e}_\beta^i$,
\begin{equation}
\label{eq:Rmatrix}
    \begin{pmatrix} \hat{x} \\ \hat{y} \\ \hat{z}   \end{pmatrix} = \begin{pmatrix}  \cos\theta_i\cos\phi_i & -\sin\phi_i & \sin\theta_i\cos\phi_i \\ \cos\theta_i\sin\phi_i & \cos\phi_i & \sin\theta_i\sin\phi_i \\ -\sin\theta_i & 0 & \cos\theta_i  \end{pmatrix} \begin{pmatrix}  \hat{e}_1^i \\ \hat{e}_2^i \\ \hat{e}_3^i   \end{pmatrix}.
\end{equation}
Note that $(R^i)^{-1} = (R^i)^{T}$ since it is an SO(3) rotation matrix. Values of $\{\theta_i, \phi_i \}$ are given by the GS.

This can then be inserted in the Hamiltonian,
\begin{align}
    H_{\text{ex}}  =& -\sum_{\langle i,j \rangle} \sum_{\alpha=\{x,y,z\}} J_\alpha S_{i\alpha}S_{j\alpha} = -\sum_{\langle i,j \rangle} \sum_{\alpha=\{1,2,3\}} J_{\hat{r}_\alpha}  (\boldsymbol{S}_{i}\cdot \hat{r}_\alpha)(\boldsymbol{S}_{j}\cdot \hat{r}_\alpha) \nonumber\\
    =& -\sum_{\langle i,j \rangle} \sum_{\alpha\beta\gamma=\{1,2,3\}} J_{\hat{r}_\alpha} (\boldsymbol{S}_i \cdot \hat{e}_\beta^i ) (\boldsymbol{S}_j \cdot \hat{e}_\gamma^j ) R_{\alpha\beta}^i R_{\alpha\gamma}^j = -\sum_{\langle i,j \rangle} J_{\hat{r}_\alpha} S_{i\beta} S_{j\gamma} R_{\alpha\beta}^i R_{\alpha\gamma}^j, \nonumber \\
    H_{\text{DM}} =& \sum_{\langle i,j \rangle} \boldsymbol{D}_{ij} \cdot [\boldsymbol{S}_i \cross \boldsymbol{S}_j] = \sum_{\langle i,j \rangle}\sum_{\alpha\beta\gamma\delta\epsilon} \epsilon_{\alpha\beta\gamma} D_{ij\hat{r}_\alpha} (\boldsymbol{S}_i \cdot \hat{e}_\delta^i ) (\boldsymbol{S}_j \cdot \hat{e}_\epsilon^j ) R_{\beta\delta}^i R_{\gamma\epsilon}^j \nonumber \\
    =& \sum_{\langle i,j \rangle} \epsilon_{\alpha\beta\gamma} D_{ij\hat{r}_\alpha} S_{i\delta} S_{j\epsilon} R_{\beta\delta}^i R_{\gamma\epsilon}^j, \nonumber \\
    H_{\text{A}} =& - K\sum_i S_{iz}^2 = -K\sum_i \Big[\sum_\alpha(\boldsymbol{S}_i \cdot \hat{e}_\alpha^i)R_{3\alpha}^i \Big]^2 = -K\sum_i (S_{i\alpha}R_{3\alpha}^i)^2, \nonumber   \\
    H_4 =& U\sum_{ijkl}^\diamond \bqty{(\boldsymbol{S}_i \cdot \boldsymbol{S}_j)(\boldsymbol{S}_k \cdot \boldsymbol{S}_l) + (\boldsymbol{S}_i \cdot \boldsymbol{S}_l)(\boldsymbol{S}_j \cdot \boldsymbol{S}_k)-(\boldsymbol{S}_i \cdot \boldsymbol{S}_k)(\boldsymbol{S}_j \cdot \boldsymbol{S}_l)} \nonumber \\
    =& U\sum_{ijkl}^\diamond \big[ (S_{i\beta}S_{j\gamma} R_{\alpha\beta}^i R_{\alpha\gamma}^j)(S_{k\epsilon}S_{l\zeta} R_{\delta\epsilon}^k R_{\delta\zeta}^l)  +(S_{i\beta}S_{l\gamma} R_{\alpha\beta}^i R_{\alpha\gamma}^l)(S_{j\epsilon}S_{k\zeta} R_{\delta\epsilon}^j R_{\delta\zeta}^k) \nonumber \\
    & -(S_{i\beta}S_{k\gamma} R_{\alpha\beta}^i R_{\alpha\gamma}^k)(S_{j\epsilon}S_{l\zeta} R_{\delta\epsilon}^j R_{\delta\zeta}^l) \big].
\end{align}
We have defined $S_{i\alpha} \equiv \boldsymbol{S}_i \cdot \hat{e}_\alpha^i$ for $\alpha = \{1,2,3\} $ and adopted the Einstein summation convention over Greek letters running over \{1,2,3\}. For the sake of generality, we keep the possibility of anisotropic exchange open here.

The HP transformation is introduced as $\boldsymbol{S}_i \cdot \boldsymbol{m}_i = S - a_i^\dagger a_i$, $S_{i\pm} = \boldsymbol{S}_i \cdot \hat{e}_1^i \pm i \boldsymbol{S}_i \cdot \hat{e}_2^i$, $S_{i+} = \sqrt{2S-a_i^\dagger a_i}~a_i, S_{i-} = a_i^\dagger\sqrt{2S-a_i^\dagger a_i}$. Here, $a_i^\dagger$ ($a_i$) creates (destroys) a magnon at lattice site $i$. We truncate at second order in magnon operators from now on, an approximation which should be valid at low temperature compared to the magnon gap \cite{QSkOP}.
$S_{i\pm} = S_{i1} \pm i S_{i2}$ leads to
\begin{equation}
    S_{i1} = \frac{1}{2}(S_{i+}+S_{i-}) = \sqrt{\frac{S}{2}}(a_i + a_i^\dagger), \qquad S_{i2} = \frac{1}{2i}(S_{i+}-S_{i-}) = i\sqrt{\frac{S}{2}}(a_i^\dagger - a_i).
\end{equation}
Finally, $S_{i3} = S - a_i^\dagger a_i$ and we are ready to introduce the HP transformation. We consider one term of the Hamiltonian at a time, and write out the sums over indices in $S_i$. For $H_{\text{ex}}$ we write out the sums over $\beta$ and $\gamma$. Separating into operator-independent terms, linear terms and quadratic terms yields
\begin{align}
    \label{eq:HexHP}
    H_{\text{ex},0} &= -S^2\sum_{\langle ij \rangle} J_{\hat{r}_\alpha}R_{\alpha3}^{i} R_{\alpha3}^{j},  \nonumber \\
    H_{\text{ex},1} &= -S\sqrt{2S} \sum_{\langle ij \rangle} [J_{\hat{r}_\alpha}(R_{\alpha1}^{i} R_{\alpha3}^{j} -iR_{\alpha2}^{i} R_{\alpha3}^{j} )a_i + \text{H.c.}], \nonumber \\
    H_{\text{ex},2} &= -\frac{S}{2} \sum_{\langle ij \rangle} \big[ -4J_{\hat{r}_\alpha}R_{\alpha3}^{i} R_{\alpha3}^{j} a_i^\dagger a_i \nonumber \\
    &\mbox{\qquad\qquad} +J_{\hat{r}_\alpha}(R_{\alpha1}^{i} R_{\alpha1}^{j} - R_{\alpha2}^{i} R_{\alpha2}^{j} -iR_{\alpha1}^{i} R_{\alpha2}^{j} -iR_{\alpha2}^{i} R_{\alpha1}^{j})a_i a_j + \text{H.c.} \nonumber\\
    &\mbox{\qquad\qquad} +J_{\hat{r}_\alpha}(R_{\alpha1}^{i} R_{\alpha1}^{j} + R_{\alpha2}^{i} R_{\alpha2}^{j} +iR_{\alpha1}^{i} R_{\alpha2}^{j} -iR_{\alpha2}^{i} R_{\alpha1}^{j})a_i a_j^\dagger + \text{H.c.} \big]. 
\end{align}
Here, we performed some rewrites so that, e.g., $H_{\text{ex},1}$ only depends on operators at lattice site $i$. H.c.~denotes the Hermitian conjugate of the preceding term. Specializing to isotropic exchange, defining $\hat{e}_\pm^i = \hat{e}_1^i \pm i \hat{e}_2^i$ and using that the columns of $R^i$ are the unit vectors $\hat{e}_\alpha^i$ we can simplify,
\begin{align}
    \label{eq:Hex2HP}
    H_{\text{ex},2} &= -\frac{JS}{2} \sum_{\langle ij \rangle} \big[ -4\hat{e}_3^{i}\cdot \hat{e}_3^j a_i^\dagger a_i +\hat{e}_-^i \cdot \hat{e}_-^j a_i a_j + \text{H.c.} +\hat{e}_-^i \cdot \hat{e}_+^j a_i a_j^\dagger + \text{H.c.} \big]. 
\end{align}

For $H_{\text{DM}}$ we write out the sums over $\delta$ and $\epsilon$ and get
\begin{align}
    \label{eq:HDMHP1st}
    H_{\text{DM},0} =& S^2 \sum_{\langle ij \rangle} \epsilon_{\alpha\beta\gamma} D_{ij\hat{r}_\alpha} R_{\beta3}^{i} R_{\gamma3}^{j}, \nonumber \\
    H_{\text{DM},1} =& S\sqrt{\frac{S}{2}} \sum_{\langle ij \rangle} \epsilon_{\alpha\beta\gamma} D_{ij\hat{r}_\alpha} [(R_{\beta1}^{i} R_{\gamma3}^{j}-iR_{\beta2}^{i} R_{\gamma3}^{j})a_i + \text{H.c.} \nonumber \\
    & + (R_{\beta3}^{i} R_{\gamma1}^{j}-iR_{\beta3}^{i} R_{\gamma2}^{j})a_j + \text{H.c.}], \nonumber \\
    H_{\text{DM},2} =& \frac{S}{2} \sum_{\langle ij \rangle} \epsilon_{\alpha\beta\gamma} D_{ij\hat{r}_\alpha} [-2R_{\beta3}^{i} R_{\gamma3}^{j}(a_i^\dagger a_i + a_j^\dagger a_j) \nonumber \\
    & + (R_{\beta1}^{i} R_{\gamma1}^{j}-R_{\beta2}^{i} R_{\gamma2}^{j}-iR_{\beta1}^{i} R_{\gamma2}^{j}-iR_{\beta2}^{i} R_{\gamma1}^{j})a_i a_j + \text{H.c.} \nonumber \\
    & + (R_{\beta1}^{i} R_{\gamma1}^{j}+R_{\beta2}^{i} R_{\gamma2}^{j}+iR_{\beta1}^{i} R_{\gamma2}^{j}-iR_{\beta2}^{i} R_{\gamma1}^{j})a_i a_j^\dagger + \text{H.c.}].
\end{align}
We seek to rewrite this to a slightly simpler form, but need to be more careful than for the exchange interaction. In $H_{\text{DM},1}$ we let $i \leftrightarrow j$ in the term with $a_j$. This is ok for a sum $\sum_{\langle ij \rangle}$. Then we rename $\beta \leftrightarrow \gamma$ and use $\epsilon_{\alpha\gamma\beta} = -\epsilon_{\alpha\beta\gamma}$ and $D_{ji\hat{r}_\alpha} = -D_{ij\hat{r}_\alpha}$ to find that the $a_j$ term can be rewritten to be equivalent to the $a_i$ term. 
Similar rewrites also show that the $a_j^\dagger a_j$ terms can be rewritten as the $a_i^\dagger a_i$ terms. Hence,
\begin{align}
    \label{eq:HDMHP}
    H_{\text{DM},1} =& S\sqrt{2S} \sum_{\langle ij \rangle} [\epsilon_{\alpha\beta\gamma} D_{ij\hat{r}_\alpha} (R_{\beta1}^{i} R_{\gamma3}^{j}-iR_{\beta2}^{i} R_{\gamma3}^{j})a_i + \text{H.c.}], \nonumber \\
    H_{\text{DM},2} =& \frac{S}{2} \sum_{\langle ij \rangle} \boldsymbol{D}_{ij} \cdot [-4 (\hat{e}_3^i \cross \hat{e}_3^j) a_i^\dagger a_i  + (\hat{e}_-^i \cross \hat{e}_-^j)a_i a_j + \text{H.c.}+ (\hat{e}_-^i \cross \hat{e}_+^j) a_i a_j^\dagger + \text{H.c.}].
\end{align}

In $H_{\text{A}}$, we perform the sum over $\alpha$, then square the sum and replace all $S_{i\alpha}S_{i\alpha'}$ by their HP transformation. This gives
\begin{align}
    H_{\text{A},0} =& -KS^2\sum_i (R_{33}^{i})^2 = -KS^2 \sum_i \cos^2\theta_i, \nonumber \\
    H_{\text{A},1} =& -KS\sqrt{2S}\sum_i [(R_{31}^{i}R_{33}^{i}-iR_{32}^{i}R_{33}^{i})a_i + \text{H.c.}] = KS\sqrt{2S}\sum_i (\sin\theta_i\cos\theta_i a_i + \text{H.c.}), \nonumber \\
    H_{\text{A},2} =& -K\frac{S}{2}\sum_i \big([-4(R_{33}^{i})^2+(R_{31}^{i})^2+(R_{32}^{i})^2]a_i^\dagger a_i + [(R_{31}^{i})^2+(R_{32}^{i})^2] a_i a_i^\dagger \nonumber \\
    &+[(R_{31}^{i})^2-(R_{32}^{i})^2-2iR_{31}^{i}R_{32}^{i}]a_i a_i + \text{H.c.}\big) \nonumber \\
    =& -\frac{KS}{2}\sum_i [(\sin^2\theta_i-4\cos^2\theta_i)a_i^\dagger a_i + \sin^2\theta_i a_i a_i^\dagger + \sin^2\theta_i a_i a_i + \text{H.c.}].
\end{align}
We inserted the definition of the rotation matrix $R^{i}$ from Eq.~\eqref{eq:Rmatrix} in order to clean up the expressions. The fact that $R_{32}^{i}=0$ allows significant simplification.

In $H_4$ we perform the sums over $\beta, \gamma, \epsilon$ and $\zeta$. Then, we need only keep the terms involving at least two $S_{i' 3}$ from all possible $S_{i\beta}S_{j\gamma}S_{k\epsilon}S_{l\zeta}$. All other terms will be more than quadratic in magnon operators. We focus on the first term $ijkl$ when performing the HP transformation. The second, $iljk$, and third, $-ikjl$, terms can then be obtained by permuting $ijkl$ appropriately. 
\begin{equation}
    H_{4,0} = US^4 \sum_{ijkl}^\diamond (R_{\alpha3}^{i} R_{\alpha3}^{j}R_{\delta3}^{k} R_{\delta3}^{l} + R_{\alpha3}^{i} R_{\alpha3}^{l}R_{\delta3}^{j} R_{\delta3}^{k} -R_{\alpha3}^{i} R_{\alpha3}^{k}R_{\delta3}^{j} R_{\delta3}^{l}).
\end{equation}
The linear terms originating from the $ijkl$ term are
\begin{align}
    [ijkl]_1 = US^3\sqrt{\frac{S}{2}}\sum_{ijkl}^\diamond [&(R_{\alpha1}^{i} R_{\alpha3}^{j}R_ {\delta3}^{k} R_{\delta3}^{l}-iR_{\alpha2}^{i} R_{\alpha3}^{j}R_{\delta3}^{k} R_{\delta3}^{l})a_i + \text{H.c.} \nonumber \\
    &+(R_{\alpha3}^{i} R_{\alpha1}^{j}R_ {\delta3}^{k} R_{\delta3}^{l}-iR_{\alpha3}^{i} R_{\alpha2}^{j}R_{\delta3}^{k} R_{\delta3}^{l})a_j + \text{H.c.} \nonumber \\
    &+(R_{\alpha3}^{i} R_{\alpha3}^{j}R_ {\delta1}^{k} R_{\delta3}^{l}-iR_{\alpha3}^{i} R_{\alpha3}^{j}R_{\delta2}^{k} R_{\delta3}^{l})a_k + \text{H.c.} \nonumber \\
    &+(R_{\alpha3}^{i} R_{\alpha3}^{j}R_ {\delta3}^{k} R_{\delta1}^{l}-iR_{\alpha3}^{i} R_{\alpha3}^{j}R_{\delta3}^{k} R_{\delta2}^{l})a_l + \text{H.c.}].
\end{align}
The best way to treat the linear terms is to rewrite all of them to $a_i$ form and then check that the total coefficient in front of $a_i$ is zero. All these terms originate from $(\boldsymbol{S}_i \cdot \boldsymbol{S}_j)(\boldsymbol{S}_k \cdot \boldsymbol{S}_l)$. We can clearly interchange $i \leftrightarrow j$ and/or $k \leftrightarrow l$ without changing this expression. We can also interchange $(i,j) \leftrightarrow (k,l)$ since $(\boldsymbol{S}_i \cdot \boldsymbol{S}_j)(\boldsymbol{S}_k \cdot \boldsymbol{S}_l) = (\boldsymbol{S}_k \cdot \boldsymbol{S}_l)(\boldsymbol{S}_i \cdot \boldsymbol{S}_j)$. Hence, in the second line of $[ijkl]_1$ we can let $i \leftrightarrow j$ to see that it is the same as the first line. In the third line we can let $(i,j) \leftrightarrow (k,l)$ to see that it is the same as the first line. In the fourth line we let $(i,j) \leftrightarrow (k,l)$ and then $i \leftrightarrow j$ to see that it is the same as the first line. In total therefore, we simply have 4 times the first line:
\begin{equation}
    [ijkl]_1 = 2US^3\sqrt{2S}\sum_{ijkl}^\diamond [(R_{\alpha1}^{i} R_{\alpha3}^{j}R_ {\delta3}^{k} R_{\delta3}^{l}-iR_{\alpha2}^{i} R_{\alpha3}^{j}R_{\delta3}^{k} R_{\delta3}^{l})a_i + \text{H.c.}]
\end{equation}
Permuting to get the $iljk$ and $-ikjl$ terms gives
\begin{align}
    H_{4,1} = 2US^3\sqrt{2S}&\sum_{ijkl}^\diamond [(R_{\alpha1}^{i} R_{\alpha3}^{j}R_ {\delta3}^{k} R_{\delta3}^{l}+R_{\alpha1}^{i} R_{\alpha3}^{l}R_ {\delta3}^{j} R_{\delta3}^{k}-R_{\alpha1}^{i} R_{\alpha3}^{k}R_ {\delta3}^{j} R_{\delta3}^{l} \nonumber \\
    &-iR_{\alpha2}^{i} R_{\alpha3}^{j}R_{\delta3}^{k} R_{\delta3}^{l} -iR_{\alpha2}^{i} R_{\alpha3}^{l}R_{\delta3}^{j} R_{\delta3}^{k}+iR_{\alpha2}^{i} R_{\alpha3}^{k}R_{\delta3}^{j} R_{\delta3}^{l})a_i + \text{H.c.}].
\end{align}
We take a similar approach to the quadratic terms. From $ijkl$ we get
\begin{align}
    &[ijkl]_2 = \frac{US^3}{2}\sum_{ijkl}^\diamond [-2 R_{\alpha3}^{i} R_{\alpha3}^{j}R_{\delta3}^{k} R_{\delta3}^{l} (a_i^\dagger a_i + a_j^\dagger a_j + a_k^\dagger a_k + a_l^\dagger a_l) \nonumber \\
    &+(R_{\alpha1}^{i} R_{\alpha1}^{j}R_{\delta3}^{k} R_{\delta3}^{l}-R_{\alpha2}^{i} R_{\alpha2}^{j}R_{\delta3}^{k} R_{\delta3}^{l}-iR_{\alpha1}^{i} R_{\alpha2}^{j}R_{\delta3}^{k} R_{\delta3}^{l}-iR_{\alpha2}^{i} R_{\alpha1}^{j}R_{\delta3}^{k} R_{\delta3}^{l}) a_i a_j + \text{H.c.} \nonumber \\
    &+(R_{\alpha1}^{i} R_{\alpha1}^{j}R_{\delta3}^{k} R_{\delta3}^{l}+R_{\alpha2}^{i} R_{\alpha2}^{j}R_{\delta3}^{k} R_{\delta3}^{l}+iR_{\alpha1}^{i} R_{\alpha2}^{j}R_{\delta3}^{k} R_{\delta3}^{l}-iR_{\alpha2}^{i} R_{\alpha1}^{j}R_{\delta3}^{k} R_{\delta3}^{l}) a_i a_j^\dagger + \text{H.c.} \nonumber \\
    &+(R_{\alpha1}^{i} R_{\alpha3}^{j}R_{\delta1}^{k} R_{\delta3}^{l}-R_{\alpha2}^{i} R_{\alpha3}^{j}R_{\delta2}^{k} R_{\delta3}^{l}-iR_{\alpha1}^{i} R_{\alpha3}^{j}R_{\delta2}^{k} R_{\delta3}^{l}-iR_{\alpha2}^{i} R_{\alpha3}^{j}R_{\delta1}^{k} R_{\delta3}^{l}) a_i a_k + \text{H.c.} \nonumber \\
    &+(R_{\alpha1}^{i} R_{\alpha3}^{j}R_{\delta1}^{k} R_{\delta3}^{l}+R_{\alpha2}^{i} R_{\alpha3}^{j}R_{\delta2}^{k} R_{\delta3}^{l}+iR_{\alpha1}^{i} R_{\alpha3}^{j}R_{\delta2}^{k} R_{\delta3}^{l}-iR_{\alpha2}^{i} R_{\alpha3}^{j}R_{\delta1}^{k} R_{\delta3}^{l}) a_i a_k^\dagger + \text{H.c.} \nonumber \\
    &+(R_{\alpha1}^{i} R_{\alpha3}^{j}R_{\delta3}^{k} R_{\delta1}^{l}-R_{\alpha2}^{i} R_{\alpha3}^{j}R_{\delta3}^{k} R_{\delta2}^{l}-iR_{\alpha1}^{i} R_{\alpha3}^{j}R_{\delta3}^{k} R_{\delta2}^{l}-iR_{\alpha2}^{i} R_{\alpha3}^{j}R_{\delta3}^{k} R_{\delta1}^{l}) a_i a_l + \text{H.c.} \nonumber \\
    &+(R_{\alpha1}^{i} R_{\alpha3}^{j}R_{\delta3}^{k} R_{\delta1}^{l}+R_{\alpha2}^{i} R_{\alpha3}^{j}R_{\delta3}^{k} R_{\delta2}^{l}+iR_{\alpha1}^{i} R_{\alpha3}^{j}R_{\delta3}^{k} R_{\delta2}^{l}-iR_{\alpha2}^{i} R_{\alpha3}^{j}R_{\delta3}^{k} R_{\delta1}^{l}) a_i a_l^\dagger + \text{H.c.} \nonumber \\
    &+(R_{\alpha3}^{i} R_{\alpha1}^{j}R_{\delta1}^{k} R_{\delta3}^{l}-R_{\alpha3}^{i} R_{\alpha2}^{j}R_{\delta2}^{k} R_{\delta3}^{l}-iR_{\alpha3}^{i} R_{\alpha1}^{j}R_{\delta2}^{k} R_{\delta3}^{l}-iR_{\alpha3}^{i} R_{\alpha2}^{j}R_{\delta1}^{k} R_{\delta3}^{l}) a_j a_k + \text{H.c.} \nonumber \\
    &+(R_{\alpha3}^{i} R_{\alpha1}^{j}R_{\delta1}^{k} R_{\delta3}^{l}+R_{\alpha3}^{i} R_{\alpha2}^{j}R_{\delta2}^{k} R_{\delta3}^{l}+iR_{\alpha3}^{i} R_{\alpha1}^{j}R_{\delta2}^{k} R_{\delta3}^{l}-iR_{\alpha3}^{i} R_{\alpha2}^{j}R_{\delta1}^{k} R_{\delta3}^{l}) a_j a_k^\dagger + \text{H.c.} \nonumber \\
    &+(R_{\alpha3}^{i} R_{\alpha1}^{j}R_{\delta3}^{k} R_{\delta1}^{l}-R_{\alpha3}^{i} R_{\alpha2}^{j}R_{\delta3}^{k} R_{\delta2}^{l}-iR_{\alpha3}^{i} R_{\alpha1}^{j}R_{\delta3}^{k} R_{\delta2}^{l}-iR_{\alpha3}^{i} R_{\alpha2}^{j}R_{\delta3}^{k} R_{\delta1}^{l}) a_j a_l + \text{H.c.} \nonumber \\
    &+(R_{\alpha3}^{i} R_{\alpha1}^{j}R_{\delta3}^{k} R_{\delta1}^{l}+R_{\alpha3}^{i} R_{\alpha2}^{j}R_{\delta3}^{k} R_{\delta2}^{l}+iR_{\alpha3}^{i} R_{\alpha1}^{j}R_{\delta3}^{k} R_{\delta2}^{l}-iR_{\alpha3}^{i} R_{\alpha2}^{j}R_{\delta3}^{k} R_{\delta1}^{l}) a_j a_l^\dagger + \text{H.c.} \nonumber \\
    &+(R_{\alpha3}^{i} R_{\alpha3}^{j}R_{\delta1}^{k} R_{\delta1}^{l}-R_{\alpha3}^{i} R_{\alpha3}^{j}R_{\delta2}^{k} R_{\delta2}^{l}-iR_{\alpha3}^{i} R_{\alpha3}^{j}R_{\delta1}^{k} R_{\delta2}^{l}-iR_{\alpha3}^{i} R_{\alpha3}^{j}R_{\delta2}^{k} R_{\delta1}^{l}) a_k a_l + \text{H.c.} \nonumber \\
    &+(R_{\alpha3}^{i} R_{\alpha3}^{j}R_{\delta1}^{k} R_{\delta1}^{l}+R_{\alpha3}^{i} R_{\alpha3}^{j}R_{\delta2}^{k} R_{\delta2}^{l}+iR_{\alpha3}^{i} R_{\alpha3}^{j}R_{\delta1}^{k} R_{\delta2}^{l}-iR_{\alpha3}^{i} R_{\alpha3}^{j}R_{\delta2}^{k} R_{\delta1}^{l}) a_k a_l^\dagger + \text{H.c.}].
\end{align}
We attempt to simplify this by making more of the sum centered at site $i$, instead of, e.g., counting contributions from only sites $k$ and $l$ here. For the $a_j^\dagger a_j$ term we let $i \leftrightarrow j$. For $a_k^\dagger a_k$ we let $(i,j) \leftrightarrow (k,l)$ and then rename $\alpha \leftrightarrow \delta$. For $a_l^\dagger a_l$ we let $(i,j) \leftrightarrow (k,l)$, then $i \leftrightarrow j$, and finally rename $\alpha \leftrightarrow \delta$. Then, this is just 4 times the $a_i^\dagger a_i$ term. We can bring the $ a_k a_l$ and $ a_k a_l^\dagger$ terms on the same form as the $a_i a_j$ and $a_i a_j^\dagger$ terms by interchanging $(i,j) \leftrightarrow (k,l)$ and then rename $\alpha \leftrightarrow \delta$. We can bring the $ a_j a_l$ and $ a_j a_l^\dagger$ terms on the same form as the $a_i a_k$ and $a_i a_k^\dagger$ terms by interchanging $i \leftrightarrow j$ and $k \leftrightarrow l$. We can bring the $ a_j a_k$ and $ a_j a_k^\dagger$ terms on the same form as the $a_i a_l$ and $a_i a_l^\dagger$ terms by interchanging $i \leftrightarrow j$ and $k \leftrightarrow l$. 

In total, we therefore have
\begin{align}
    &[ijkl]_2 = US^3\sum_{ijkl}^\diamond [-4 R_{\alpha3}^{i} R_{\alpha3}^{j}R_{\delta3}^{k} R_{\delta3}^{l} a_i^\dagger a_i  \nonumber \\
    &+(R_{\alpha1}^{i} R_{\alpha1}^{j}R_{\delta3}^{k} R_{\delta3}^{l}-R_{\alpha2}^{i} R_{\alpha2}^{j}R_{\delta3}^{k} R_{\delta3}^{l}-iR_{\alpha1}^{i} R_{\alpha2}^{j}R_{\delta3}^{k} R_{\delta3}^{l}-iR_{\alpha2}^{i} R_{\alpha1}^{j}R_{\delta3}^{k} R_{\delta3}^{l}) a_i a_j + \text{H.c.} \nonumber \\
    &+(R_{\alpha1}^{i} R_{\alpha1}^{j}R_{\delta3}^{k} R_{\delta3}^{l}+R_{\alpha2}^{i} R_{\alpha2}^{j}R_{\delta3}^{k} R_{\delta3}^{l}+iR_{\alpha1}^{i} R_{\alpha2}^{j}R_{\delta3}^{k} R_{\delta3}^{l}-iR_{\alpha2}^{i} R_{\alpha1}^{j}R_{\delta3}^{k} R_{\delta3}^{l}) a_i a_j^\dagger + \text{H.c.} \nonumber \\
    &+(R_{\alpha1}^{i} R_{\alpha3}^{j}R_{\delta1}^{k} R_{\delta3}^{l}-R_{\alpha2}^{i} R_{\alpha3}^{j}R_{\delta2}^{k} R_{\delta3}^{l}-iR_{\alpha1}^{i} R_{\alpha3}^{j}R_{\delta2}^{k} R_{\delta3}^{l}-iR_{\alpha2}^{i} R_{\alpha3}^{j}R_{\delta1}^{k} R_{\delta3}^{l}) a_i a_k + \text{H.c.} \nonumber \\
    &+(R_{\alpha1}^{i} R_{\alpha3}^{j}R_{\delta1}^{k} R_{\delta3}^{l}+R_{\alpha2}^{i} R_{\alpha3}^{j}R_{\delta2}^{k} R_{\delta3}^{l}+iR_{\alpha1}^{i} R_{\alpha3}^{j}R_{\delta2}^{k} R_{\delta3}^{l}-iR_{\alpha2}^{i} R_{\alpha3}^{j}R_{\delta1}^{k} R_{\delta3}^{l}) a_i a_k^\dagger + \text{H.c.} \nonumber \\
    &+(R_{\alpha1}^{i} R_{\alpha3}^{j}R_{\delta3}^{k} R_{\delta1}^{l}-R_{\alpha2}^{i} R_{\alpha3}^{j}R_{\delta3}^{k} R_{\delta2}^{l}-iR_{\alpha1}^{i} R_{\alpha3}^{j}R_{\delta3}^{k} R_{\delta2}^{l}-iR_{\alpha2}^{i} R_{\alpha3}^{j}R_{\delta3}^{k} R_{\delta1}^{l}) a_i a_l + \text{H.c.} \nonumber \\
    &+(R_{\alpha1}^{i} R_{\alpha3}^{j}R_{\delta3}^{k} R_{\delta1}^{l}+R_{\alpha2}^{i} R_{\alpha3}^{j}R_{\delta3}^{k} R_{\delta2}^{l}+iR_{\alpha1}^{i} R_{\alpha3}^{j}R_{\delta3}^{k} R_{\delta2}^{l}-iR_{\alpha2}^{i} R_{\alpha3}^{j}R_{\delta3}^{k} R_{\delta1}^{l}) a_i a_l^\dagger + \text{H.c.}].
\end{align}
Permuting to get the $iljk$ and $-ikjl$ terms gives
\begin{align}
    H_{4,2} &= US^3\sum_{ijkl}^\diamond \big(-4 [(\hat{e}_3^i \cdot \hat{e}_3^j)(\hat{e}_3^k \cdot \hat{e}_3^l) + (\hat{e}_3^i \cdot \hat{e}_3^l)(\hat{e}_3^j \cdot \hat{e}_3^k) - (\hat{e}_3^i \cdot \hat{e}_3^j)(\hat{e}_3^k \cdot \hat{e}_3^l)] a_i^\dagger a_i  \nonumber \\
    &+[(\hat{e}_-^i \cdot \hat{e}_-^j)(\hat{e}_3^k \cdot \hat{e}_3^l) + (\hat{e}_-^i \cdot \hat{e}_3^l)(\hat{e}_-^j \cdot \hat{e}_3^k) - (\hat{e}_-^i \cdot \hat{e}_3^k)(\hat{e}_-^j \cdot \hat{e}_3^l)] a_i a_j + \text{H.c.} \nonumber \\
    &+[(\hat{e}_-^i \cdot \hat{e}_+^j)(\hat{e}_3^k \cdot \hat{e}_3^l) + (\hat{e}_-^i \cdot \hat{e}_3^l)(\hat{e}_+^j \cdot \hat{e}_3^k) - (\hat{e}_-^i \cdot \hat{e}_3^k)(\hat{e}_+^j \cdot \hat{e}_3^l)] a_i a_j^\dagger + \text{H.c.} \nonumber \\
    &+[(\hat{e}_-^i \cdot \hat{e}_3^j)(\hat{e}_-^k \cdot \hat{e}_3^l) + (\hat{e}_-^i \cdot \hat{e}_3^l)(\hat{e}_3^j \cdot \hat{e}_-^k) - (\hat{e}_-^i \cdot \hat{e}_-^k)(\hat{e}_3^j \cdot \hat{e}_3^l)] a_i a_k + \text{H.c.} \nonumber \\
    &+[(\hat{e}_-^i \cdot \hat{e}_3^j)(\hat{e}_+^k \cdot \hat{e}_3^l) + (\hat{e}_-^i \cdot \hat{e}_3^l)(\hat{e}_3^j \cdot \hat{e}_+^k) - (\hat{e}_-^i \cdot \hat{e}_+^k)(\hat{e}_3^j \cdot \hat{e}_3^l)] a_i a_k^\dagger + \text{H.c.} \nonumber \\
    &+[(\hat{e}_-^i \cdot \hat{e}_3^j)(\hat{e}_3^k \cdot \hat{e}_-^l) + (\hat{e}_-^i \cdot \hat{e}_-^l)(\hat{e}_3^j \cdot \hat{e}_3^k) - (\hat{e}_-^i \cdot \hat{e}_3^k)(\hat{e}_3^j \cdot \hat{e}_-^l)] a_i a_l + \text{H.c.} \nonumber \\
    &+[(\hat{e}_-^i \cdot \hat{e}_3^j)(\hat{e}_3^k \cdot \hat{e}_+^l) + (\hat{e}_-^i \cdot \hat{e}_+^l)(\hat{e}_3^j \cdot \hat{e}_3^k) - (\hat{e}_-^i \cdot \hat{e}_3^k)(\hat{e}_3^j \cdot \hat{e}_+^l)] a_i a_l^\dagger + \text{H.c.}\big).
\end{align}

\subsection{Fourier transform}
The Fourier transform (FT) is introduced as
\begin{equation}
    \label{eq:FT}
    a_i = \frac{1}{\sqrt{N^{(r)}}} \sum_{\boldsymbol{k}}^{(r)} e^{i\boldsymbol{k}\cdot \boldsymbol{r}_i} a_{\boldsymbol{k}}^{(r)},
\end{equation}
\begin{equation}
    \label{eq:FTinv}
    a_{\boldsymbol{k}}^{(r)} = \frac{1}{\sqrt{N^{(r)}}} \sum_{i}^{(r)} e^{-i\boldsymbol{k}\cdot \boldsymbol{r}_i}a_i .
\end{equation}
We assume lattice site $i$, located at $\boldsymbol{r}_i$, resides on sublattice $r$. $N^{(r)}$ is the number of lattice sites on sublattice $r$, the sum over $\boldsymbol{k}$ is restricted to the first Brillouin zone (1BZ) of sublattice $r$, 1BZ$^{(r)}$, and $a_{\boldsymbol{k}}^{(r)}$ is a magnon destruction operator associated with sublattice $r$. 

The sum in the exchange Hamiltonian is converted in the following way
\begin{equation}
    \sum_{\langle ij \rangle} = \sum_{\langle r s \rangle} \sum_i^{(r)} \sum_{\boldsymbol{\delta}_{(r,s)}}.
\end{equation}
Here, we sum over all sublattices $r$ and for each $r$ we sum over those sublattices $s$ that contain sites which are nearest neighbors to a site on sublattice $r$. We sum over all sites $i$ on sublattice $r$ and all possible vectors $\boldsymbol{\delta}_{(r,s)}$ connecting two sites $i$ and $j$ that are nearest neighbors, located on sublattice $r$ and $s$, respectively.

An example of a FT of a term in $H_{\text{ex}}$
\begin{align}
    \sum_{\langle ij \rangle} \hat{e}_-^i \cdot \hat{e}_-^j a_i a_j  =\sum_{\langle rs \rangle} \hat{e}_-^r \cdot \hat{e}_-^s \sum_{i}^{(r)} \sum_{\boldsymbol{\delta}_{(r,s)}}\frac{1}{\sqrt{N^{(r)}}} \sum_{\boldsymbol{k}}^{(r)} e^{i\boldsymbol{k}\cdot \boldsymbol{r}_i} a_{\boldsymbol{k}}^{(r)}  \frac{1}{\sqrt{N^{(s)}}} \sum_{\boldsymbol{k}'}^{(s)} e^{i\boldsymbol{k}'\cdot (\boldsymbol{r}_i+\boldsymbol{\delta}_{(r,s)})} a_{\boldsymbol{k}'}^{(s)}.
\end{align}
Since the orthonormal frame $\{\hat{e}_1^i, \hat{e}_2^i, \hat{e}_3^i\}$ is the same for all lattice sites $i$ on sublattice $r$, we can move $\hat{e}_\alpha^{i}$ outside the sum over $i$ by writing $\hat{e}_\alpha^r$. Now focus on the FT:
\begin{align}
    \frac{1}{\sqrt{N^{(r)}N^{(s)}}}& \sum_{\boldsymbol{k}}^{(r)}\sum_{\boldsymbol{k}'}^{(s)} \sum_{i}^{(r)} e^{i(\boldsymbol{k}+\boldsymbol{k}')\cdot \boldsymbol{r}_i}\sum_{\boldsymbol{\delta}_{(r,s)}}e^{i\boldsymbol{k}'\cdot \boldsymbol{\delta}_{(r,s)}} a_{\boldsymbol{k}}^{(r)}   a_{\boldsymbol{k}'}^{(s)}  =\sqrt{\frac{N^{(r)}}{N^{(s)}}} \sum_{\boldsymbol{k}}^{(r)}\sum_{\boldsymbol{k}'}^{(s)} \delta_{\boldsymbol{k},-\boldsymbol{k}'} \sum_{\boldsymbol{\delta}_{(r,s)}}e^{i\boldsymbol{k}'\cdot \boldsymbol{\delta}_{(r,s)}} a_{\boldsymbol{k}}^{(r)}   a_{\boldsymbol{k}'}^{(s)}  \nonumber \\
    &=\sqrt{\frac{N^{(r)}}{N^{(s)}}} \sum_{\boldsymbol{k}}^{(r, -s)} \Big(\sum_{\boldsymbol{\delta}_{(r,s)}}e^{-i\boldsymbol{k}\cdot \boldsymbol{\delta}_{(r,s)}}\Big) a_{\boldsymbol{k}}^{(r)}   a_{-\boldsymbol{k}}^{(s)} = \sqrt{\frac{N^{(r)}}{N^{(s)}}} \sum_{\boldsymbol{k}}^{(r, -s)} \gamma_{\boldsymbol{k}}^{(r,s)} a_{\boldsymbol{k}}^{(r)}   a_{-\boldsymbol{k}}^{(s)}.
\end{align}
We used that $\sum_{i}^{(r)} e^{i(\boldsymbol{k}+\boldsymbol{k}')\cdot \boldsymbol{r}_i} = N^{(r)}\delta_{\boldsymbol{k},-\boldsymbol{k'}}$. It is clear that in the end we only get contributions from those $\boldsymbol{k}$ such that $\boldsymbol{k} \in \text{1BZ}^{(r)}$ and $-\boldsymbol{k} \in \text{1BZ}^{(s)}$. This is indicated by $(r,-s)$ above the summation sign. We also introduce a quantity $z_{rs}^{\text{NN}} = \gamma_{\boldsymbol{0}}^{(r,s)}$ which is the number of lattice sites $j$ on a specific sublattice $s$ that are nearest neighbors (NN) to lattice site $i$ on sublattice $r$. Following this approach, the end result is that
\begin{align}
    H_{\text{ex},2} = \sum_{\langle rs \rangle} \bigg(& F_{\text{ex},rdr}^{(r,s)} \sum_{\boldsymbol{k}}^{(r)} a_{\boldsymbol{k}}^{(r)\dagger} a_{\boldsymbol{k}}^{(r)} +F_{\text{ex},rs}^{(r,s)} \sum_{\boldsymbol{k}}^{(r, -s)} \gamma_{\boldsymbol{k}}^{(r,s)} a_{\boldsymbol{k}}^{(r)}   a_{-\boldsymbol{k}}^{(s)} + \text{H.c.} \nonumber \\
    &+F_{\text{ex},rsd}^{(r,s)}  \sum_{\boldsymbol{k}}^{(r, s)} \gamma_{\boldsymbol{k}}^{(r,s)} a_{\boldsymbol{k}}^{(r)}   a_{\boldsymbol{k}}^{(s)\dagger} + \text{H.c.} \bigg), 
\end{align}
where we defined
\begin{align}
    F_{\text{ex},rdr}^{(r,s)} &= 2JS \hat{e}_3^r \cdot \hat{e}_3^s  z_{rs}^{\text{NN}}, \\
    F_{\text{ex},rs}^{(r,s)} &= -\frac{JS}{2}\sqrt{\frac{N^{(r)}}{N^{(s)}}}  \hat{e}_-^r \cdot \hat{e}_-^s, \\
    F_{\text{ex},rsd}^{(r,s)} &= -\frac{JS}{2} \sqrt{\frac{N^{(r)}}{N^{(s)}}} \hat{e}_-^r \cdot \hat{e}_+^s . 
\end{align}
Subscript $r,s$ denote which kind of sublattice combination the following operators belong to, and $d$ is added to indicate a creation operator instead of an annihilation operator. Superscript $rs$ indicates that the prefactor depends on the specific sublattices. 

For the DMI terms, we need to take into account that the DMI vector depends on the direction of the vector connecting $i$ and $j$, namely $\boldsymbol{\delta}_{(r,s)}$. Hence, $D_{ij\hat{r}_\alpha} \to D_{(r,s) \hat{r}_\alpha}$ when we rewrite the sum. Note that $\boldsymbol{\delta}_{(r,s)}$ is the shortest vector connecting two lattice sites on sublattice $r$ and $s$. 
Then, we find
\begin{align}
    H_{\text{DM},2} = \sum_{\langle rs \rangle} \bigg(&F_{\text{DM},rdr}^{(r,s)} \sum_{\boldsymbol{k}}^{(r)}   a_{\boldsymbol{k}}^{(r)\dagger} a_{\boldsymbol{k}}^{(r)}  + F_{\text{DM},rs}^{(r,s)} \sum_{\boldsymbol{k}}^{(r, -s)} \gamma_{\boldsymbol{k}}^{(r,s)} a_{\boldsymbol{k}}^{(r)}   a_{-\boldsymbol{k}}^{(s)} + \text{H.c.} \nonumber \\
    & + F_{\text{DM},rsd}^{(r,s)} \sum_{\boldsymbol{k}}^{(r, s)} \gamma_{\boldsymbol{k}}^{(r,s)} a_{\boldsymbol{k}}^{(r)}   a_{\boldsymbol{k}}^{(s)\dagger} + \text{H.c.}\bigg),
\end{align}
with
\begin{align}
    F_{\text{DM},rdr}^{(r,s)} &= -2S\boldsymbol{D}_{(r,s)} \cdot (\hat{e}_3^r \cross \hat{e}_3^s)z_{rs}^{\text{NN}},\\
    F_{\text{DM},rs}^{(r,s)} &=  \frac{S}{2}\sqrt{\frac{N^{(r)}}{N^{(s)}}}\boldsymbol{D}_{(r,s)} \cdot (\hat{e}_-^r \cross \hat{e}_-^s) , \\
    F_{\text{DM},rsd}^{(r,s)}&=  \frac{S}{2}\sqrt{\frac{N^{(r)}}{N^{(s)}}}\boldsymbol{D}_{(r,s)} \cdot (\hat{e}_-^r \cross \hat{e}_+^s).
\end{align}

For $H_{\text{A},2}$ we get
\begin{align}
    H_{\text{A},2} = \sum_r \bigg( \sum_{\boldsymbol{k}}^{(r)}[ &F_{\text{A},rdr}^{(r)} a_{\boldsymbol{k}}^{(r)\dagger} a_{\boldsymbol{k}}^{(r)} + F_{\text{A},rr}^{(r)}  a_{\boldsymbol{k}}^{(r)} a_{\boldsymbol{k}}^{(r)\dagger}] +\sum_{\boldsymbol{k}}^{(r,-r)} [ F_{\text{A},rr}^{(r)} a_{\boldsymbol{k}}^{(r)}   a_{-\boldsymbol{k}}^{(r)} + \text{H.c.}] \bigg),
\end{align}
\begin{align}
    F_{\text{A},rdr}^{(r)} = -\frac{KS}{2}(\sin^2 \theta_r - 4\cos^2\theta_r) \mbox{\quad and \quad} F_{\text{A},rr}^{(r)} = -\frac{KS}{2}\sin^2 \theta_r
\end{align}

For $H_4$ we rewrite the sum as
\begin{equation}
    \sum_{ijkl}^\diamond \to \sum_{rstu}^\diamond \sum_i^{(r)}\sum_{\boldsymbol{\delta}_{(r,s)}}\sum_{\boldsymbol{\delta}_{(r,t)}}\sum_{\boldsymbol{\delta}_{(r,u)}}
\end{equation}
We sum over all sublattices $r$. Then, for each $r$ we sum over sublattices $s,t,u$ such that sites $i,j,k,l$ on sublattices $r,s,t,u$ form diamonds oriented counterclockwise of minimal area \cite{HeinzeSkX, H4PRB}. Then we sum over all sites $i$ on sublattice $r$ and all vectors connecting $i$ to sites $j,k,l$ such that $i,j,k,l$ form counterclockwise diamonds and where $j,k,l$ are located on sublattices $s,t,u$. Let us introduce
$z_{rs}^{\diamond}$ as the number of sites on sublattice $s$ that can form diamonds with a specific site on sublattice $r$. Often this is one, but let us be general. Then, we find that
\begin{align}
    H_{4,2} =& \sum_{rstu}^\diamond \bigg[F_{4,rdr}^{(r,s,t,u)} \sum_{\boldsymbol{k}}^{(r)} a_{\boldsymbol{k}}^{(r)\dagger} a_{\boldsymbol{k}}^{(r)}  \nonumber \\
    &+F_{4,rs}^{(r,s,t,u)} \sum_{\boldsymbol{k}}^{(r, -s)} \gamma_{\boldsymbol{k}}^{(r,s)} a_{\boldsymbol{k}}^{(r)}   a_{-\boldsymbol{k}}^{(s)} + \text{H.c.} +F_{4,rsd}^{(r,s,t,u)}  \sum_{\boldsymbol{k}}^{(r, s)} \gamma_{\boldsymbol{k}}^{(r,s)} a_{\boldsymbol{k}}^{(r)}   a_{\boldsymbol{k}}^{(s)\dagger}  + \text{H.c.} \nonumber \\
    &+F_{4,rt}^{(r,s,t,u)} \sum_{\boldsymbol{k}}^{(r, -t)} \gamma_{\boldsymbol{k}}^{(r,t)} a_{\boldsymbol{k}}^{(r)}   a_{-\boldsymbol{k}}^{(t)} + \text{H.c.}+F_{4,rtd}^{(r,s,t,u)}  \sum_{\boldsymbol{k}}^{(r, t)} \gamma_{\boldsymbol{k}}^{(r,t)} a_{\boldsymbol{k}}^{(r)}   a_{\boldsymbol{k}}^{(t)\dagger}  + \text{H.c.} \nonumber \\
    &+F_{4,ru}^{(r,s,t,u)}  \sum_{\boldsymbol{k}}^{(r, -u)} \gamma_{\boldsymbol{k}}^{(r,u)} a_{\boldsymbol{k}}^{(r)}   a_{-\boldsymbol{k}}^{(u)} + \text{H.c.} +F_{4,rud}^{(r,s,t,u)}  \sum_{\boldsymbol{k}}^{(r, u)} \gamma_{\boldsymbol{k}}^{(r,u)} a_{\boldsymbol{k}}^{(r)}   a_{\boldsymbol{k}}^{(u)\dagger}  + \text{H.c.}\bigg],
\end{align}
with
\begin{align}
    F_{4,rdr}^{(r,s,t,u)} =& -4US^3 z_{rs}^\diamond z_{rt}^\diamond z_{ru}^\diamond[(\hat{e}_3^r \cdot \hat{e}_3^s)(\hat{e}_3^t \cdot \hat{e}_3^u) + (\hat{e}_3^r \cdot \hat{e}_3^u)(\hat{e}_3^s \cdot \hat{e}_3^t) - (\hat{e}_3^r \cdot \hat{e}_3^t)(\hat{e}_3^s \cdot \hat{e}_3^u)] , \\
    F_{4,rs}^{(r,s,t,u)} =& US^3 z_{rt}^\diamond z_{ru}^\diamond \sqrt{\frac{N^{(r)}}{N^{(s)}}}[(\hat{e}_-^r \cdot \hat{e}_-^s)(\hat{e}_3^t \cdot \hat{e}_3^u) + (\hat{e}_-^r \cdot \hat{e}_3^u)(\hat{e}_-^s \cdot \hat{e}_3^t) - (\hat{e}_-^r \cdot \hat{e}_3^t)(\hat{e}_-^s \cdot \hat{e}_3^u)],  \\
    F_{4,rsd}^{(r,s,t,u)} =& US^3 z_{rt}^\diamond z_{ru}^\diamond \sqrt{\frac{N^{(r)}}{N^{(s)}}}[(\hat{e}_-^r \cdot \hat{e}_+^s)(\hat{e}_3^t \cdot \hat{e}_3^u) + (\hat{e}_-^r \cdot \hat{e}_3^u)(\hat{e}_+^s \cdot \hat{e}_3^t) - (\hat{e}_-^r \cdot \hat{e}_3^t)(\hat{e}_+^s \cdot \hat{e}_3^u)] , \\
    F_{4,rt}^{(r,s,t,u)} =&US^3 z_{rs}^\diamond z_{ru}^\diamond  \sqrt{\frac{N^{(r)}}{N^{(t)}}} [(\hat{e}_-^r \cdot \hat{e}_3^s)(\hat{e}_-^t \cdot \hat{e}_3^u) + (\hat{e}_-^r \cdot \hat{e}_3^u)(\hat{e}_3^s \cdot \hat{e}_-^t) - (\hat{e}_-^r \cdot \hat{e}_-^t)(\hat{e}_3^s \cdot \hat{e}_3^u)] ,\\
    F_{4,rtd}^{(r,s,t,u)} =& US^3 z_{rs}^\diamond z_{ru}^\diamond \sqrt{\frac{N^{(r)}}{N^{(t)}}}[(\hat{e}_-^r \cdot \hat{e}_3^s)(\hat{e}_+^t \cdot \hat{e}_3^u) + (\hat{e}_-^r \cdot \hat{e}_3^u)(\hat{e}_3^s \cdot \hat{e}_+^t) - (\hat{e}_-^r \cdot \hat{e}_+^t)(\hat{e}_3^s \cdot \hat{e}_3^u)] ,\\
    F_{4,ru}^{(r,s,t,u)} =& US^3 z_{rs}^\diamond z_{rt}^\diamond \sqrt{\frac{N^{(r)}}{N^{(u)}}} [(\hat{e}_-^r \cdot \hat{e}_3^s)(\hat{e}_3^t \cdot \hat{e}_-^u) + (\hat{e}_-^r \cdot \hat{e}_-^u)(\hat{e}_3^s \cdot \hat{e}_3^t) - (\hat{e}_-^r \cdot \hat{e}_3^t)(\hat{e}_3^s \cdot \hat{e}_-^u)] ,\\
    F_{4,rud}^{(r,s,t,u)} =& US^3 z_{rs}^\diamond z_{rt}^\diamond \sqrt{\frac{N^{(r)}}{N^{(u)}}} [(\hat{e}_-^r \cdot \hat{e}_3^s)(\hat{e}_3^t \cdot \hat{e}_+^u) + (\hat{e}_-^r \cdot \hat{e}_+^u)(\hat{e}_3^s \cdot \hat{e}_3^t) - (\hat{e}_-^r \cdot \hat{e}_3^t)(\hat{e}_3^s \cdot \hat{e}_+^u)]  .
\end{align}

\subsection{Collecting the total Hamiltonian}
Collecting all terms we get 
\begin{align}
    H_0 =& -JS^2\sum_{\langle ij \rangle}  R_{\alpha3}^{i} R_{\alpha3}^{j}  +S^2 \sum_{\langle ij \rangle} \epsilon_{\alpha\beta\gamma} D_{ij\hat{r}_\alpha} R_{\beta3}^{i} R_{\gamma3}^{j} -KS^2\sum_i  \cos^2\theta_i \nonumber \\
    &+US^4 \sum_{ijkl}^\diamond ( R_{\alpha3}^{i} R_{\alpha3}^{j}R_{\delta3}^{k} R_{\delta3}^{l} + R_{\alpha3}^{i} R_{\alpha3}^{l}R_{\delta3}^{j} R_{\delta3}^{k} -R_{\alpha3}^{i} R_{\alpha3}^{k}R_{\delta3}^{j} R_{\delta3}^{l} ).
\end{align}
This is simply the classical Hamiltonian from Eq.~\eqref{eq:Hclass}.

All linear terms in the Hamiltonian should vanish if we expand around the correct ground state of the system \cite{HProtation_2009}. Collecting, we find
\begin{align}
\label{eq:H1gen}
    H_1 =& S\sqrt{2S}\sum_i \bigg[ \sum_{j \in \text{NN}} \big[-J(R_{\alpha1}^{i} R_{\alpha3}^{j} -iR_{\alpha2}^{i} R_{\alpha3}^{j} )+\epsilon_{\alpha\beta\gamma} D_{ij\hat{r}_\alpha} (R_{\beta1}^{i} R_{\gamma3}^{j}-iR_{\beta2}^{i} R_{\gamma3}^{j})\big] +K \sin\theta_i\cos\theta_i \nonumber \\
    &+2US^2\sum_{jkl}^\diamond \big(R_{\alpha1}^{i} R_{\alpha3}^{j}R_ {\delta3}^{k} R_{\delta3}^{l}+R_{\alpha1}^{i} R_{\alpha3}^{l}R_ {\delta3}^{j} R_{\delta3}^{k}-R_{\alpha1}^{i} R_{\alpha3}^{k}R_ {\delta3}^{j} R_{\delta3}^{l} \nonumber \\
    &-iR_{\alpha2}^{i} R_{\alpha3}^{j}R_{\delta3}^{k} R_{\delta3}^{l} -iR_{\alpha2}^{i} R_{\alpha3}^{l}R_{\delta3}^{j} R_{\delta3}^{k}+iR_{\alpha2}^{i} R_{\alpha3}^{k}R_{\delta3}^{j} R_{\delta3}^{l}\big) \bigg] a_i + \text{H.c.}
\end{align}
We sum over all $j$ that are nearest neighbors (NNs) to $i$, and all $jkl$ that can form counterclockwise diamonds of minimal area with $i$.
The linear terms are zero if the quantity inside the large square brackets is zero at all $i$.

Collecting the FTed quadratic part gives
\begin{align}
\label{eq:H2gen}
    H_{2} &=  \sum_{\langle rs \rangle} \sum_{\boldsymbol{k}}^{(r)}  [ F_{\text{ex},rdr}^{(r,s)} + F_{\text{DM},rdr}^{(r,s)}] a_{\boldsymbol{k}}^{(r)\dagger} a_{\boldsymbol{k}}^{(r)}+ \sum_{rstu}^\diamond  \sum_{\boldsymbol{k}}^{(r)}F_{4,rdr}^{(r,s,t,u)} a_{\boldsymbol{k}}^{(r)\dagger} a_{\boldsymbol{k}}^{(r)}  \nonumber \\
    &+\bigg(\sum_{\langle rs \rangle}\sum_{\boldsymbol{k}}^{(r, -s)} [F_{\text{ex},rs}^{(r,s)} +  F_{\text{DM},rs}^{(r,s)}] \gamma_{\boldsymbol{k}}^{(r,s)} a_{\boldsymbol{k}}^{(r)}   a_{-\boldsymbol{k}}^{(s)}  + \sum_{rstu}^\diamond \sum_{\boldsymbol{k}}^{(r, -s)} F_{4,rs}^{(r,s,t,u)}\gamma_{\boldsymbol{k}}^{(r,s)} a_{\boldsymbol{k}}^{(r)}   a_{-\boldsymbol{k}}^{(s)}  \nonumber \\
    &+  \sum_{\langle rs \rangle} \sum_{\boldsymbol{k}}^{(r, s)}[F_{\text{ex},rsd}^{(r,s)} +  F_{\text{DM},rsd}^{(r,s)}]\gamma_{\boldsymbol{k}}^{(r,s)} a_{\boldsymbol{k}}^{(r)}   a_{\boldsymbol{k}}^{(s)\dagger}  + \sum_{rstu}^\diamond \sum_{\boldsymbol{k}}^{(r, s)} F_{4,rsd}^{(r,s,t,u)}\gamma_{\boldsymbol{k}}^{(r,s)} a_{\boldsymbol{k}}^{(r)}   a_{\boldsymbol{k}}^{(s)\dagger} \bigg) + \text{H.c.} \nonumber \\
    &+\sum_r \bigg( \sum_{\boldsymbol{k}}^{(r)} F_{\text{A},rdr}^{(r)} a_{\boldsymbol{k}}^{(r)\dagger} a_{\boldsymbol{k}}^{(r)} + \sum_{\boldsymbol{k}}^{(r)}F_{\text{A},rr}^{(r)}  a_{\boldsymbol{k}}^{(r)} a_{\boldsymbol{k}}^{(r)\dagger} +\sum_{\boldsymbol{k}}^{(r,-r)} [ F_{\text{A},rr}^{(r)} a_{\boldsymbol{k}}^{(r)}   a_{-\boldsymbol{k}}^{(r)} + \text{H.c.}] \bigg) \nonumber \\
    &+\sum_{rstu}^\diamond \bigg(  \sum_{\boldsymbol{k}}^{(r, -t)} F_{4,rt}^{(r,s,t,u)} \gamma_{\boldsymbol{k}}^{(r,t)} a_{\boldsymbol{k}}^{(r)}   a_{-\boldsymbol{k}}^{(t)} + \sum_{\boldsymbol{k}}^{(r, t)} F_{4,rtd}^{(r,s,t,u)} \gamma_{\boldsymbol{k}}^{(r,t)} a_{\boldsymbol{k}}^{(r)}   a_{\boldsymbol{k}}^{(t)\dagger}   \nonumber \\
    &\mbox{\qquad}+  \sum_{\boldsymbol{k}}^{(r, -u)} F_{4,ru}^{(r,s,t,u)}  \gamma_{\boldsymbol{k}}^{(r,u)} a_{\boldsymbol{k}}^{(r)}   a_{-\boldsymbol{k}}^{(u)}  +  \sum_{\boldsymbol{k}}^{(r, u)}F_{4,rud}^{(r,s,t,u)} \gamma_{\boldsymbol{k}}^{(r,u)} a_{\boldsymbol{k}}^{(r)}   a_{\boldsymbol{k}}^{(u)\dagger}\bigg)  + \text{H.c.}
\end{align}

\subsection{Analytic proof that linear terms vanish} \label{sec:linzero}
Here, we include an analytic proof that the terms that are linear in magnon operators are zero when expanding around the GS of the system. Checking numerically that the linear terms in Eq.~\eqref{eq:H1gen} are zero therefore serves as a check on whether we have obtained the true GS from the numerical simulations. In Appendix A2 of Ref.~\cite{QSkOP} we argue that the linear terms are zero within numerical accuracy.

Picking a specific lattice site $i'$ we can require that $H_1 = 0$ by setting the real and imaginary parts of the coefficients to zero,
\begin{align}
    &\sum_{j' \in \text{NN}} \big(-J_{\hat{r}_\alpha}R_{\alpha1}^{i'} R_{\alpha3}^{j'} +\epsilon_{\alpha\beta\gamma} D_{i'j'\hat{r}_\alpha} R_{\beta1}^{i'} R_{\gamma3}^{j'}\big) +K \sin\theta_{i'}\cos\theta_{i'} \nonumber \\
    &+2US^2\sum_{j'k'l'}^\diamond \big(R_{\alpha1}^{i'} R_{\alpha3}^{j'}R_ {\delta3}^{k'} R_{\delta3}^{l'}+R_{\alpha1}^{i'} R_{\alpha3}^{l'}R_ {\delta3}^{j'} R_{\delta3}^{k'}-R_{\alpha1}^{i'} R_{\alpha3}^{k'}R_ {\delta3}^{j'} R_{\delta3}^{l'} \big) = 0, \label{eq:reH1=0} \\
    &\sum_{j' \in \text{NN}} \big(-J_{\hat{r}_\alpha}R_{\alpha2}^{i'} R_{\alpha3}^{j'} +\epsilon_{\alpha\beta\gamma} D_{i'j'\hat{r}_\alpha} R_{\beta2}^{i'} R_{\gamma3}^{j'}\big)  \nonumber \\
    &+2US^2\sum_{j'k'l'}^\diamond \big(R_{\alpha2}^{i'} R_{\alpha3}^{j'}R_{\delta3}^{k'} R_{\delta3}^{l'} +R_{\alpha2}^{i'} R_{\alpha3}^{l'}R_{\delta3}^{j'} R_{\delta3}^{k'} - R_{\alpha2}^{i'} R_{\alpha3}^{k'}R_{\delta3}^{j'} R_{\delta3}^{l'}\big) = 0. \label{eq:imH1=0}
\end{align}

We now show that setting $\pdv{H_0}{\theta_{i'}} = 0$ and $\pdv{H_0}{\phi_{i'}} = 0$ leads to the same constraints. It is important to take into account that any $ijkl$ in the sum could be $i'$. If, e.g., $j = i'$, we cyclically permute so that $k = j'$, $l = k'$ and $i = l'$. Then we see that all choices lead to the same sums, and so we can simply write, e.g., 4 times the case where $i = i', j = j', k=k', l=l'$. We also use that, from the definition in Eq.~\eqref{eq:Rmatrix},
\begin{align}
    \pdv{R_{\alpha3}^{i'}}{\theta_{i'}} &= R_{\alpha1}^{i'}, \mbox{\quad and \quad }\pdv{R_{\alpha3}^{i'}}{\phi_{i'}} = \sin\theta_{i'}R_{\alpha2}^{i'}.
\end{align}
So we get
\begin{align}
    \pdv{H_0}{\theta_{i'}} =& 2S^2 \bigg( \sum_{j' \in \text{NN}} \big(-J_{\hat{r}_\alpha}R_{\alpha1}^{i'} R_{\alpha3}^{j'} +\epsilon_{\alpha\beta\gamma} D_{i'j'\hat{r}_\alpha} R_{\beta1}^{i'} R_{\gamma3}^{j'}\big) +K \sin\theta_{i'}\cos\theta_{i'} \nonumber \\
    &+2US^2\sum_{j'k'l'}^\diamond \big(R_{\alpha1}^{i'} R_{\alpha3}^{j'}R_ {\delta3}^{k'} R_{\delta3}^{l'}+R_{\alpha1}^{i'} R_{\alpha3}^{l'}R_ {\delta3}^{j'} R_{\delta3}^{k'}-R_{\alpha1}^{i'} R_{\alpha3}^{k'}R_ {\delta3}^{j'} R_{\delta3}^{l'} \big)  \bigg) ,\\
    \pdv{H_0}{\phi_{i'}} =& 2S^2 \sin\theta_{i'} \bigg( \sum_{j' \in \text{NN}} \big(-J_{\hat{r}_\alpha}R_{\alpha2}^{i'} R_{\alpha3}^{j'} +\epsilon_{\alpha\beta\gamma} D_{i'j'\hat{r}_\alpha} R_{\beta2}^{i'} R_{\gamma3}^{j'}\big)  \nonumber \\
    &+2US^2\sum_{j'k'l'}^\diamond \big(R_{\alpha2}^{i'} R_{\alpha3}^{j'}R_{\delta3}^{k'} R_{\delta3}^{l'} +R_{\alpha2}^{i'} R_{\alpha3}^{l'}R_{\delta3}^{j'} R_{\delta3}^{k'} - R_{\alpha2}^{i'} R_{\alpha3}^{k'}R_{\delta3}^{j'} R_{\delta3}^{l'}\big) \bigg).
\end{align}
The case $\sin\theta_{i'} = 0$ for all $i'$ is uninteresting in this context, so we see that requiring $\pdv{H_0}{\theta_{i'}} = 0$ and $\pdv{H_0}{\phi_{i'}} = 0$ leads to the same constraints as we had for $H_1 = 0$ in Eqs.~\eqref{eq:reH1=0} and \eqref{eq:imH1=0}.

Hence, if we are in an extremum of $H_0$, e.g.~the GS, $H_1 = 0$. This is also fairly obvious from arguments presented in Ref.~\cite{HProtation_2009}; the function $H(\{a_i, a_i^\dagger\})$ should be in a minimum with respect to the operators, and so all linear terms must vanish. 

\subsection{Specializing to the ground states}

\begin{figure}
    \centering
    \includegraphics[width = 0.6\textwidth]{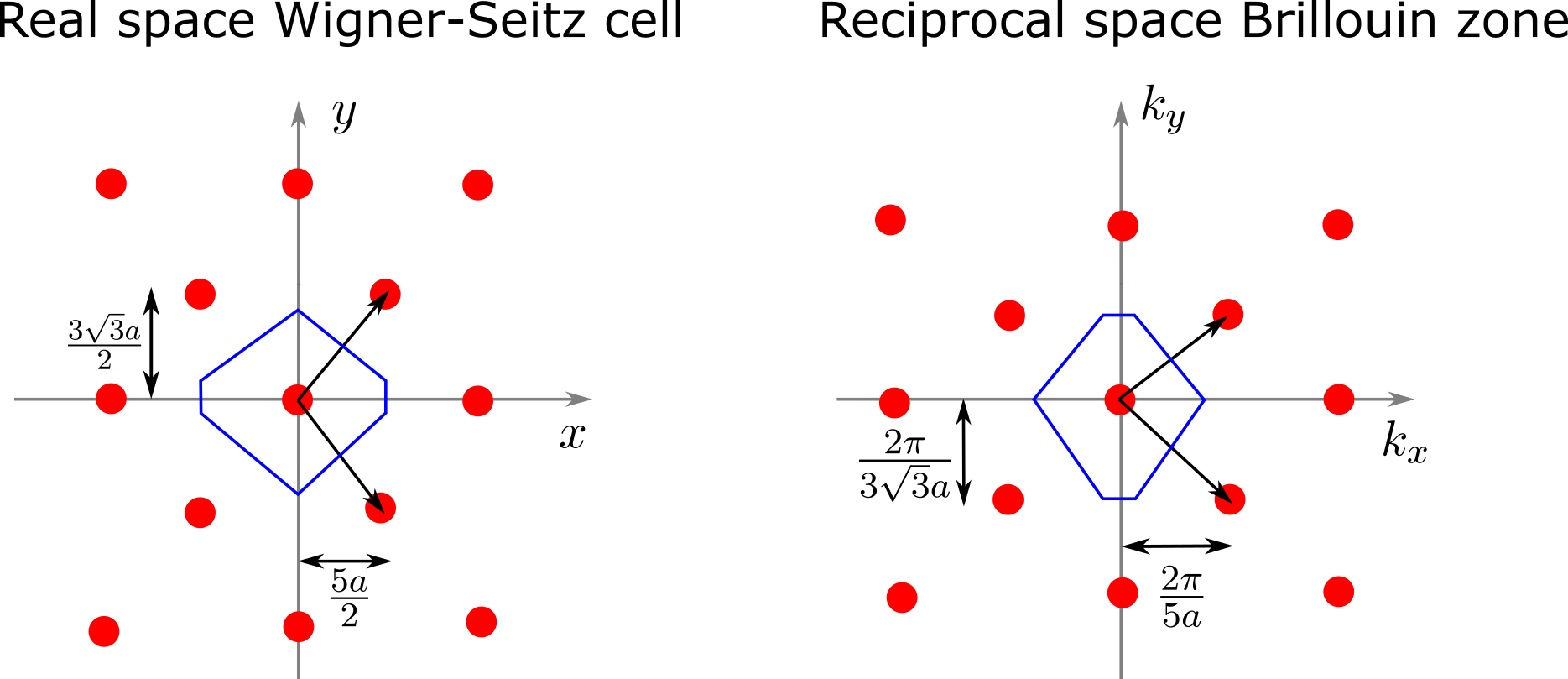}
    \caption{Wigner-Seitz cell and first Brilloin zone for the sublattices. The difference between lattice spacing in the $x$ and $y$ directions has been exaggerated in the figure.}
    \label{fig:WSBZ}
\end{figure}

SkX1 and SkX2 contain 15 sublattices that are equal, centered rectangular lattices with primitive vectors $\boldsymbol{a}_1 = (5/2, -3\sqrt{3}/2)$ and $\boldsymbol{a}_2 = (5/2, 3\sqrt{3}/2)$. Hence, all sublattices have the same 1BZ and the same number of lattice sites, i.e., all $N^{(r)}$ are equal. We name them $N' = N/N_{\text{SL}}$, where $N$ is the total number of lattice sites, and $N_{\text{SL}} = 15$ is the number of sublattices. In reciprocal space the primitive vectors are
$\boldsymbol{b}_1 = 2\pi \pqty{1/5, -1/3\sqrt{3}}$ and  $\boldsymbol{b}_2 = 2\pi \pqty{1/5, 1/3\sqrt{3}}$. Hence, the 1BZ is a nonregular hexagon with corners at $(\pm 52\pi/135, 0)$, $( 2\pi/135,\pm 2\pi/3\sqrt{3})$ and $(- 2\pi/135,\pm 2\pi/3\sqrt{3})$. The 1BZ is inversion symmetric. Therefore, all sums over momentum are restricted to the same values of $\boldsymbol{k}$, and we can replace all of them with $\sum_{\boldsymbol{k}}$ with the understanding that $\boldsymbol{k}$ is restricted to the 1BZ of the sublattices. See Fig.~\ref{fig:WSBZ} for sketches of the Wigner-Seitz cell in real space and the 1BZ in reciprocal space.

Furthermore, the SkX1 and SkX2 GSs are such that we never encounter a situation where more than one site on the same sublattice is a nearest neighbor to a specific site on another sublattice, nor can make diamonds in $H_4$. Hence, all the $z_{rs}^{\text{NN}}, z_{rs}^{\diamond}$ factors are 1, and all sums over vectors connecting lattice sites only contain one term, $\gamma_{\boldsymbol{k}}^{(r,s)} = e^{-i \boldsymbol{k} \cdot \boldsymbol{\delta}_{(r,s)} }$.


To get any further we need to adopt a numbering of the sublattices, and then perform the sums over the sublattices to obtain a matrix version of the total quadratic Hamiltonian. We use the numbering shown in Fig.~\ref{fig:MWC}. With reference to that numbering we state for reference the 12 counterclockwise diamonds in $H_4$ where sublattice $r = 10$. These are $(r,s,t,u) = \{ (10, 11, 7, 6),$  $(10, 11, 6, 5),$ $(10, 6, 2, 5),$  $(10,6,5,9),$  $(10,5,4,9),$  $(10,5,9,13),$  $(10,9,3,13),$ $(10,9,13,14),$  $(10,13,15,14),$  $(10,13,14,11),$ $(10,14,1,11),$ $(10,14,11,6) \}$.

\subsection{Quadratic part}
Having specialized to our SkX GSs, the quadratic part can be written
\begin{align}
\label{eq:H2spec}
    H_{2} =&  \sum_{\boldsymbol{k}}\bigg\{ \sum_{\langle rs \rangle} \big[ (F_{\text{ex},rdr}^{(r,s)} + F_{\text{DM},rdr}^{(r,s)})a_{\boldsymbol{k}}^{(r)\dagger} a_{\boldsymbol{k}}^{(r)} + (F_{\text{ex},rs}^{(r,s)} + F_{\text{DM},rs}^{(r,s)})\gamma_{\boldsymbol{k}}^{(r,s)} a_{\boldsymbol{k}}^{(r)}    a_{-\boldsymbol{k}}^{(s)} + \text{H.c.} \nonumber \\
    &+ (F_{\text{ex},rsd}^{(r,s)}  +   F_{\text{DM},rsd}^{(r,s)})\gamma_{\boldsymbol{k}}^{(r,s)} a_{\boldsymbol{k}}^{(r)}   a_{\boldsymbol{k}}^{(s)\dagger}  + \text{H.c.} \big] \nonumber \\
    &+\sum_r \big[(F_{\text{A},rdr}^{(r)}+F_{\text{A},rr}^{(r)}) a_{\boldsymbol{k}}^{(r)\dagger} a_{\boldsymbol{k}}^{(r)} + F_{\text{A},rr}^{(r)} a_{\boldsymbol{k}}^{(r)}   a_{-\boldsymbol{k}}^{(r)} + \text{H.c.} \big] \nonumber \\
    &+\sum_{rstu}^\diamond \Big(F_{4,rdr}^{(r,s,t,u)} a_{\boldsymbol{k}}^{(r)\dagger} a_{\boldsymbol{k}}^{(r)} + F_{4,rs}^{(r,s,t,u)} \gamma_{\boldsymbol{k}}^{(r,s)} a_{\boldsymbol{k}}^{(r)}    a_{-\boldsymbol{k}}^{(s)} + \text{H.c.} + F_{4,rsd}^{(r,s,t,u)} \gamma_{\boldsymbol{k}}^{(r,s)} a_{\boldsymbol{k}}^{(r)}   a_{\boldsymbol{k}}^{(s)\dagger} + \text{H.c.} \nonumber \\
    &+F_{4,rt}^{(r,s,t,u)}\gamma_{\boldsymbol{k}}^{(r,t)} a_{\boldsymbol{k}}^{(r)}   a_{-\boldsymbol{k}}^{(t)} + \text{H.c.}+F_{4,rtd}^{(r,s,t,u)}  \gamma_{\boldsymbol{k}}^{(r,t)} a_{\boldsymbol{k}}^{(r)}   a_{\boldsymbol{k}}^{(t)\dagger}  + \text{H.c.} \nonumber \\
    &+ F_{4,ru}^{(r,s,t,u)} \gamma_{\boldsymbol{k}}^{(r,u)} a_{\boldsymbol{k}}^{(r)}   a_{-\boldsymbol{k}}^{(u)} + \text{H.c.} +F_{4,rud}^{(r,s,t,u)}   \gamma_{\boldsymbol{k}}^{(r,u)} a_{\boldsymbol{k}}^{(r)}   a_{\boldsymbol{k}}^{(u)\dagger}  + \text{H.c.}\Big)\bigg\}.
\end{align}
We used a commutator which leads to a shift of $H_0$, 
\begin{equation}
\label{eq:H'HA}
    H'_0 = H_0 +N' \sum_r F_{\text{A},rr}^{(r)}.
\end{equation}

Upon writing out the sums over sublattices, we organize the factors into the following types: $a_{\boldsymbol{k}}^{(r)\dagger} a_{\boldsymbol{k}}^{(r)} $, $a_{\boldsymbol{k}}^{(r)\dagger} a_{\boldsymbol{k}}^{(s)}$ with $r$ and $s>r$ nearest neighbors, $a_{\boldsymbol{k}}^{(r)\dagger} a_{\boldsymbol{k}}^{(t)} $ with $r$ and $t>r$ next nearest neighbors, $a_{\boldsymbol{k}}^{(r)} a_{-\boldsymbol{k}}^{(r)} $, $a_{\boldsymbol{k}}^{(r)} a_{-\boldsymbol{k}}^{(s)}$ with $r$ and $s>r$ nearest neighbors, and $a_{\boldsymbol{k}}^{(r)} a_{-\boldsymbol{k}}^{(t)} $ with $r$ and $t>r$ next nearest neighbors. These are then later rewritten using commutators and letting $\boldsymbol{k} \to -\boldsymbol{k}$ in the sum where necessary, e.g.,
\begin{equation}
    \sum_{\boldsymbol{k}}c_{\boldsymbol{k}}a_{\boldsymbol{k}}^{(r)} a_{-\boldsymbol{k}}^{(s)} = \frac{1}{2} \sum_{\boldsymbol{k}}(c_{-\boldsymbol{k}}a_{-\boldsymbol{k}}^{(r)} a_{\boldsymbol{k}}^{(s)} + c_{\boldsymbol{k}} a_{-\boldsymbol{k}}^{(s)} a_{\boldsymbol{k}}^{(r)}),
\end{equation}
where $c_{\boldsymbol{k}}$ is the coefficient in front of $a_{\boldsymbol{k}}^{(r)} a_{-\boldsymbol{k}}^{(s)}$.
Contributions to $a_{\boldsymbol{k}}^{(r)\dagger} a_{\boldsymbol{k}}^{(s)}$ with $r$ and $s>r$ nearest neighbors comes from two sources; the H.c.~of $a_{\boldsymbol{k}}^{(r)}   a_{\boldsymbol{k}}^{(s)\dagger}$ and from $a_{\boldsymbol{k}}^{(r')}   a_{\boldsymbol{k}}^{(s')\dagger}$ with $r' = s >r$, $s' = r$. We find that these two coefficients are the same, and so we store the prefactor as twice the H.c. of the coefficient of $a_{\boldsymbol{k}}^{(r)}   a_{\boldsymbol{k}}^{(s)\dagger}$.

Similarly, $a_{\boldsymbol{k}}^{(r)} a_{-\boldsymbol{k}}^{(s)}$ with $r$ and $s>r$ nearest neighbors comes from two sources; $a_{\boldsymbol{k}}^{(r)} a_{-\boldsymbol{k}}^{(s)}$ and $a_{\boldsymbol{k}}^{(r')} a_{-\boldsymbol{k}}^{(s')}$ with $r' = s>r$, $s' = r$ and $\boldsymbol{k} \to -\boldsymbol{k}$. The latter is permissible due to the sum over $\boldsymbol{k}$ over an inversion symmetric 1BZ. Again, these two coefficients are found to be the same, and are stored as twice the coefficient of $a_{\boldsymbol{k}}^{(r)} a_{-\boldsymbol{k}}^{(s)}$. Finally, similar considerations apply to the terms involving $r$ and $t$ as next-nearest neighbors. 

As presented in Ref.~\cite{QSkOP} we can write
\begin{equation}
\label{eq:H2matrix}
    H_2 = \frac{1}{2}\sum_{\boldsymbol{k}}  \boldsymbol{a}_{\boldsymbol{k}}^\dagger M_{\boldsymbol{k}}  \boldsymbol{a}_{\boldsymbol{k}},
\end{equation}
where $\boldsymbol{a}_{\boldsymbol{k}}^\dagger = (a_{\boldsymbol{k}}^{(1)\dagger}, a_{\boldsymbol{k}}^{(2)\dagger}, \dots, a_{\boldsymbol{k}}^{(15)\dagger}, a_{-\boldsymbol{k}}^{(1)}, \dots, a_{-\boldsymbol{k}}^{(15)})$ and
\begin{equation}
\label{eq:Mat}
    M_{\boldsymbol{k}} = \begin{pmatrix} \eta_{\boldsymbol{k}} & \nu_{-\boldsymbol{k}}^* \\ \nu_{\boldsymbol{k}} & \eta_{-\boldsymbol{k}}^*    \end{pmatrix},
\end{equation}
The matrix elements can be written,
\begin{align}
    \eta_{\boldsymbol{k}}^{r,s} =& \eta_r \delta_{r,s} +Se^{i\boldsymbol{k}\cdot \boldsymbol{\delta}_{(r,s)}}\Lambda_+^{r,s},
\end{align}
\begin{align}
    \eta_r =& 2S\sum_s [J_{(r,s)}\hat{e}_3^r \cdot \hat{e}_3^s -\boldsymbol{D}_{(r,s)} \cdot (\hat{e}_3^r \cross \hat{e}_3^s)] -KS[1-3(\hat{e}_3^r \cdot \hat{z})^2] \nonumber \\
    &-4S^3 \sum_{s,t,u} U_{(r,s,t,u)}[(\hat{e}_3^r \cdot \hat{e}_3^s)(\hat{e}_3^t \cdot \hat{e}_3^u) + (\hat{e}_3^r \cdot \hat{e}_3^u)(\hat{e}_3^s \cdot \hat{e}_3^t) - (\hat{e}_3^r \cdot \hat{e}_3^t)(\hat{e}_3^s \cdot \hat{e}_3^u)],
\end{align}
\begin{align}
    \nu_{\boldsymbol{k}}^{r,s} =& \nu_r \delta_{r,s} +Se^{i\boldsymbol{k}\cdot \boldsymbol{\delta}_{(r,s)}}\Lambda_-^{r,s} ,
\end{align}
\begin{align}
    \nu_r =& -KS(\hat{e}_1^r \cdot \hat{z})^2 .
\end{align}
\begin{align}
    \Lambda_\pm^{r,s} =& -J_{(r,s)}\hat{e}_{\pm}^r \cdot \hat{e}_-^s + \boldsymbol{D}_{(r,s)} \cdot (\hat{e}_{\pm}^r \cross \hat{e}_-^s) \nonumber \\
    &+2S^2\bigg(\sum_{t,u} U_{(r,s,t,u)}[(\hat{e}_{\pm}^r \cdot \hat{e}_-^s)(\hat{e}_3^t \cdot \hat{e}_3^u) + (\hat{e}_{\pm}^r \cdot \hat{e}_3^u)(\hat{e}_-^s \cdot \hat{e}_3^t) - (\hat{e}_{\pm}^r \cdot \hat{e}_3^t)(\hat{e}_-^s \cdot \hat{e}_3^u)] \nonumber \\
    &+\sum_{s',u} U_{(r,s',s,u)} [(\hat{e}_{\pm}^r \cdot \hat{e}_3^{s'})(\hat{e}_-^s \cdot \hat{e}_3^u) + (\hat{e}_{\pm}^r \cdot \hat{e}_3^u)(\hat{e}_3^{s'} \cdot \hat{e}_-^s) - (\hat{e}_{\pm}^r \cdot \hat{e}_-^s)(\hat{e}_3^{s'} \cdot \hat{e}_3^u)] \nonumber \\
    &+\sum_{s',t} U_{(r,s',t,s)} [(\hat{e}_{\pm}^r \cdot \hat{e}_3^{s'})(\hat{e}_3^t \cdot \hat{e}_-^s) + (\hat{e}_{\pm}^r \cdot \hat{e}_-^s)(\hat{e}_3^{s'} \cdot \hat{e}_3^t) - (\hat{e}_{\pm}^r \cdot \hat{e}_3^t)(\hat{e}_3^{s'} \cdot \hat{e}_-^s)] \bigg).
\end{align}
Here, $J_{(r,s)} = J$ if there exists $i\in r, j\in s$ such that $i$ and $j$ are nearest neighbors. Otherwise $J_{(r,s)} = 0$. $\boldsymbol{D}_{(r,s)} = D \boldsymbol{\delta}_{(r,s)} \cross \hat{z}$ if there exist $i\in r, j\in s$ such that $i$ and $j$ are nearest neighbors. Otherwise $\boldsymbol{D}_{(r,s)} = \boldsymbol{0}$. $U_{(r,s,t,u)} = U$ if there exist $i \in r, j \in s, k \in t, l \in u$ such that sites $i,j,k,l$ make a counterclockwise diamond of minimal area. Otherwise $U_{(r,s,t,u)} = 0$. 

\subsection{Diagonalization}
We employ the method described in Ref.~\cite{COLPA} to diagonalize the system. This is an alternative to the method described in Refs.~\cite{Tsallis, Xiao}. The first method works only if the matrix $M_{\boldsymbol{k}}$ in the Hamiltonian is positive definite, unlike the latter which is more general. Following Ref.~\cite{COLPA} the matrix $M_{\boldsymbol{k}}$ in Eq.~\eqref{eq:Mat} is a $30\cross30$ matrix as opposed to $60\cross60$ if following Ref.~\cite{Tsallis}. Hence, the first method gives 15 energy bands, while the latter originally gives 30 bands, which can be reduced to 15 bands using methods similar to those we presented in Refs.~\cite{KMmaster, PWSWBEC}. We find that both methods give the same results in the present system. In SkX2, where the excitation spectrum is not inversion symmetric it would be challenging to perform the reduction from 30 to 15 bands without the result from following Ref.~\cite{COLPA} as a guide.

The matrix form of the Hamiltonian is Hermitian, $M_{\boldsymbol{k}}^\dagger = M_{\boldsymbol{k}}$, while the submatrices obey $\eta_{\boldsymbol{k}}^\dagger = \eta_{\boldsymbol{k}}$, $\nu_{\boldsymbol{k}}^T = \nu_{-\boldsymbol{k}}$. The system is diagonalized with a transformation matrix $T_{\boldsymbol{k}}$ as follows,
\begin{equation}
    \boldsymbol{a}_{\boldsymbol{k}}^\dagger M_{\boldsymbol{k}} \boldsymbol{a}_{\boldsymbol{k}} = (\boldsymbol{a}_{\boldsymbol{k}}^\dagger T_{\boldsymbol{k}}^\dagger)[(T_{\boldsymbol{k}}^{-1})^\dagger M_{\boldsymbol{k}} T_{\boldsymbol{k}}^{-1}] (T_{\boldsymbol{k}} \boldsymbol{a}_{\boldsymbol{k}}) = \boldsymbol{b}_{\boldsymbol{k}}^\dagger D_{\boldsymbol{k}} \boldsymbol{b}_{\boldsymbol{k}},
\end{equation}
where $D_{\boldsymbol{k}}$ is diagonal. The diagonalized operator vector is $\boldsymbol{b}_{\boldsymbol{k}} = (b_{\boldsymbol{k},1},$ $b_{\boldsymbol{k},2}, \dots,$ $b_{\boldsymbol{k},15},$ $b_{-\boldsymbol{k},1}^\dagger, \dots,$ $b_{-\boldsymbol{k},15}^\dagger)^T $. The diagonalized operators retain bosonic commutation relations since the transformation matrix is paraunitary \cite{COLPA}, $T_{\boldsymbol{k}}^{-1} = \mathcal{J}T_{\boldsymbol{k}}^\dagger \mathcal{J}$. Here, $\mathcal{J}$ is a diagonal matrix whose first $15$ diagonal elements are $1$, and final $15$ diagonal elements are $-1$.

The algorithm for obtaining the transformation matrix is described in Ref.~\cite{COLPA} as follows:
\begin{enumerate}
    \item Find $K_{\boldsymbol{k}}$ from the Cholesky decomposition $M_{\boldsymbol{k}} = K_{\boldsymbol{k}}^\dagger K_{\boldsymbol{k}}$. $K_{\boldsymbol{k}}^\dagger$ is lower triangular, while $K_{\boldsymbol{k}}$ is upper triangular.
    \item Find eigenvectors $\boldsymbol{w}_{\boldsymbol{k},1}, ..., \boldsymbol{w}_{\boldsymbol{k},2m}$ and eigenvalues $E_{\boldsymbol{k},1}, ..., E_{\boldsymbol{k},2m}$ of the Hermitian $2m\cross2m$ matrix $K_{\boldsymbol{k}}\mathcal{J}K_{\boldsymbol{k}}^\dagger$. We have $E_{\boldsymbol{k}, n} = -E_{-\boldsymbol{k},n+m}$ for $n \leq m$, where $E_{\boldsymbol{k}, n} > 0$ and $E_{-\boldsymbol{k},n+m} < 0$. The $m$ positive eigenvalues are the excitation spectrum of the system \cite{COLPA}.
    \item Construct a unitary matrix $W_{\boldsymbol{k}} = [\boldsymbol{w}_{\boldsymbol{k},1}| ...|\boldsymbol{w}_{\boldsymbol{k},2m}]$ from the orthonormal eigenvectors.
    \item Construct $D_{\boldsymbol{k}} = \operatorname{diag}(E_{\boldsymbol{k},1}, ..., E_{\boldsymbol{k},m}, -E_{\boldsymbol{k},m+1}, \dots, -E_{\boldsymbol{k},2m})$ from the eigenvalues $E_{\boldsymbol{k},n}$.
    \item Calculate $T_{\boldsymbol{k}}^{-1}$ row by row from $K_{\boldsymbol{k}}T_{\boldsymbol{k}}^{-1} = W_{\boldsymbol{k}}D_{\boldsymbol{k}}^{\frac{1}{2}}$ starting at the last row since $K_{\boldsymbol{k}}$ is upper triangular. \label{item:FindT_Colpa}
\end{enumerate}
Step \ref{item:FindT_Colpa} can be performed as follows,
\begin{equation}
    (T_{\boldsymbol{k}}^{-1})_{\text{row } 2m-i} = \frac{(W_{\boldsymbol{k}}D_{\boldsymbol{k}}^{\frac{1}{2}})_{\text{row } 2m-i} - \sum_{j = 0}^{i-1}K_{\boldsymbol{k},2m-i, 2m-j}(T_{\boldsymbol{k}}^{-1})_{\text{row } 2m - j}}{K_{\boldsymbol{k},2m-i,2m-i}}.
\end{equation}

Using $T_{\boldsymbol{k}}^{-1} = \mathcal{J}T_{\boldsymbol{k}}^\dagger \mathcal{J}$ gives $T_{\boldsymbol{k}} = \mathcal{J}(T_{\boldsymbol{k}}^{-1})^\dagger \mathcal{J}$. Let
\begin{equation}
    T_{\boldsymbol{k}} = \begin{pmatrix} U_{\boldsymbol{k}} & W_{\boldsymbol{k}} \\ V_{\boldsymbol{k}} & Z_{\boldsymbol{k}}    \end{pmatrix}.
\end{equation}
From $\boldsymbol{b}_{\boldsymbol{k}} = T_{\boldsymbol{k}}\boldsymbol{a}_{\boldsymbol{k}}$, we find
\begin{align}
    b_{\boldsymbol{k},n} &= \sum_{r = 1}^{m} \bqty{U_{\boldsymbol{k},n,r} a_{\boldsymbol{k}}^{(r)} + W_{\boldsymbol{k},n,r} a_{-\boldsymbol{k}}^{(r)\dagger} }, \label{eq:bkr}  \\
    b_{-\boldsymbol{k},n}^\dagger &= \sum_{r = 1}^{m} \bqty{ V_{\boldsymbol{k},n,r} a_{\boldsymbol{k}}^{(r)} + Z_{\boldsymbol{k},n,r} a_{-\boldsymbol{k}}^{(r)\dagger} }, \label{eq:bmkrdT}\\
    b_{-\boldsymbol{k},n}^\dagger &= \sum_{r = 1}^{m} \bqty{ W_{-\boldsymbol{k},n,r}^* a_{\boldsymbol{k}}^{(r)} + U_{-\boldsymbol{k},n,r}^* a_{-\boldsymbol{k}}^{(r)\dagger} }. \label{eq:bmkrddirect}
\end{align}
Eq. \eqref{eq:bmkrdT} was found using $T_{\boldsymbol{k}}$, while Eq. \eqref{eq:bmkrddirect} was found directly from Eq.~\eqref{eq:bkr} by letting $\boldsymbol{k} \to -\boldsymbol{k}$ and taking the H.c. This shows that $W_{\boldsymbol{k}} = V_{-\boldsymbol{k}}^*$, while $Z_{\boldsymbol{k}} = U_{-\boldsymbol{k}}^*$. Hence,
\begin{equation}
\label{eq:ColpaT}
    T_{\boldsymbol{k}} = \begin{pmatrix} U_{\boldsymbol{k}} & V_{-\boldsymbol{k}}^* \\ V_{\boldsymbol{k}} & U_{-\boldsymbol{k}}^*    \end{pmatrix}.
\end{equation}
Using $T_{\boldsymbol{k}}^{-1} = \mathcal{J}T_{\boldsymbol{k}}^\dagger \mathcal{J}$ gives
\begin{equation}
    T_{\boldsymbol{k}}^{-1} = \begin{pmatrix} U_{\boldsymbol{k}}^\dagger & -V_{\boldsymbol{k}}^\dagger \\ -V_{-\boldsymbol{k}}^T & U_{-\boldsymbol{k}}^T    \end{pmatrix}.
\end{equation}

Diagonalizing $H_2$ in Eq.~\eqref{eq:H2matrix} yields \cite{COLPA}
\begin{equation}
    H_2 = \frac{1}{2}\sum_{\boldsymbol{k}} \sum_{n = 1}^{15} \pqty{E_{\boldsymbol{k},n}b_{\boldsymbol{k},n}^\dagger b_{\boldsymbol{k},n} - E_{\boldsymbol{k},n+m}b_{-\boldsymbol{k},n}b_{-\boldsymbol{k},n}^\dagger},
\end{equation}
Letting $-\boldsymbol{k} \to \boldsymbol{k}$ in last term of the sum gives
\begin{equation}
    H_2 = \frac{1}{2}\sum_{\boldsymbol{k}} \sum_{n = 1}^{15} \pqty{E_{\boldsymbol{k},n}b_{\boldsymbol{k},n}^\dagger b_{\boldsymbol{k},n} - \underbrace{E_{-\boldsymbol{k},n+m}}_{=-E_{\boldsymbol{k},n} }b_{\boldsymbol{k},n}b_{\boldsymbol{k},n}^\dagger},
\end{equation}
Using a commutator in the final term gives
\begin{equation}
    H_2 = \sum_{\boldsymbol{k}}\sum_{n = 1}^{15} E_{\boldsymbol{k}, n} \pqty{b_{\boldsymbol{k},n}^\dagger b_{\boldsymbol{k},n} + \frac{1}{2}}.
\end{equation}
The excitation spectrum is shown in Fig.~1 of the main text, and discussed further in Ref.~\cite{QSkOP}.

\subsubsection{Proofs}
\paragraph{Statement.}The diagonalized operator vector is
\begin{equation}
    \boldsymbol{b}_{\boldsymbol{k}} = (b_{\boldsymbol{k},1}, b_{\boldsymbol{k},2}, \dots, b_{\boldsymbol{k},15}, b_{-\boldsymbol{k},1}^\dagger, \dots, b_{-\boldsymbol{k},15}^\dagger)^T .
\end{equation}
\paragraph*{Proof.} Our original basis is
\begin{equation}
    \boldsymbol{a}_{\boldsymbol{k}} = (a_{\boldsymbol{k}}^{(1)}, a_{\boldsymbol{k}}^{(2)}, \dots, a_{\boldsymbol{k}}^{(15)}, a_{-\boldsymbol{k}}^{(1)^\dagger}, \dots, a_{-\boldsymbol{k}}^{(15)\dagger})^T.
\end{equation}
Let us define
\begin{equation}
    \Sigma_x = \begin{pmatrix} 0 & I \\ I&0    \end{pmatrix},
\end{equation}
and note that $\Sigma_x^2 = I$, where $I$ is the identity matrix. We notice that $((\Sigma_x\boldsymbol{a}_{-\boldsymbol{k}})^T)^\dagger = \boldsymbol{a}_{\boldsymbol{k}}$. We now prove that the transformation $\boldsymbol{b}_{\boldsymbol{k}} = T_{\boldsymbol{k}}\boldsymbol{a}_{\boldsymbol{k}}$ preserves this:
\begin{equation}
    \Sigma_x\boldsymbol{b}_{-\boldsymbol{k}} = \Sigma_x T_{-\boldsymbol{k}} \boldsymbol{a}_{-\boldsymbol{k}} = \Sigma_x T_{-\boldsymbol{k}} \Sigma_x \Sigma_x \boldsymbol{a}_{-\boldsymbol{k}}.
\end{equation}
Taking the transpose yields
\begin{equation}
    (\Sigma_x\boldsymbol{b}_{-\boldsymbol{k}})^T = (\Sigma_x\boldsymbol{a}_{-\boldsymbol{k}})^T \Sigma_x T_{-\boldsymbol{k}}^T \Sigma_x.
\end{equation}
The Hermitian conjugate of this is
\begin{equation}
    ((\Sigma_x\boldsymbol{b}_{-\boldsymbol{k}})^T)^\dagger = \Sigma_x T_{-\boldsymbol{k}}^* \Sigma_x ((\Sigma_x\boldsymbol{a}_{-\boldsymbol{k}})^T)^\dagger .
\end{equation}
Here, one can easily show that $\Sigma_x T_{-\boldsymbol{k}}^* \Sigma_x = T_{\boldsymbol{k}}$ using Eq. \eqref{eq:ColpaT}, and we have $((\Sigma_x\boldsymbol{a}_{-\boldsymbol{k}})^T)^\dagger = \boldsymbol{a}_{\boldsymbol{k}}$. Thus we arrive at
\begin{equation}
    ((\Sigma_x\boldsymbol{b}_{-\boldsymbol{k}})^T)^\dagger = T_{\boldsymbol{k}} \boldsymbol{a}_{\boldsymbol{k}} = \boldsymbol{b}_{\boldsymbol{k}}.
\end{equation} 
This proves that $\boldsymbol{b}_{\boldsymbol{k}}$ has the same form as $\boldsymbol{a}_{\boldsymbol{k}}$.

\paragraph*{Note.} $K_{\boldsymbol{k}}\mathcal{J}K_{\boldsymbol{k}}^\dagger$ and $M_{\boldsymbol{k}}\mathcal{J}$ are similar matrices (connected by similarity transformation, $M_{\boldsymbol{k}}\mathcal{J} = K_{\boldsymbol{k}}^\dagger K_{\boldsymbol{k}} \mathcal{J}$, so $(K_{\boldsymbol{k}}^\dagger)^{-1}M_{\boldsymbol{k}}\mathcal{J} K_{\boldsymbol{k}}^\dagger = K_{\boldsymbol{k}}\mathcal{J}K_{\boldsymbol{k}}^\dagger$). They therefore have the same eigenvalues.

\paragraph{Statement.} If $\lambda_{\boldsymbol{k}}$ is an eigenvalue of $M_{\boldsymbol{k}}J$, then $-\lambda_{\boldsymbol{k}}$ is an eigenvalue of $M_{-\boldsymbol{k}}J$. 

\paragraph*{Proof.} Introduce an operator $F_c$ such that
\begin{equation}
    F_c \begin{pmatrix} u \\ v \end{pmatrix} = \Sigma_x \begin{pmatrix} u \\ v \end{pmatrix}^* = \begin{pmatrix} v^* \\ u^* \end{pmatrix}.
\end{equation}
It can be shown that $\{\mathcal{J}, F_c\} = 0$ and $F_c M_{\boldsymbol{k}} \boldsymbol{x} = M_{-\boldsymbol{k}} F_c \boldsymbol{x}$. Thus, if $M_{\boldsymbol{k}}\mathcal{J}\boldsymbol{x} = \lambda_{\boldsymbol{k}} \boldsymbol{x}$,
\begin{equation}
    M_{-\boldsymbol{k}}\mathcal{J}F_c \boldsymbol{x} = -F_c M_{\boldsymbol{k}}\mathcal{J}\boldsymbol{x} = -F_c \lambda_{\boldsymbol{k}}\boldsymbol{x} =  -\lambda_{\boldsymbol{k}} F_c\boldsymbol{x},
\end{equation}
which shows that if $\boldsymbol{x}$ is an eigenvector of $M_{\boldsymbol{k}}\mathcal{J}$ with eigenvalue $\lambda_{\boldsymbol{k}} \in \mathbb{R}$, then $F_c\boldsymbol{x}$ is an eigenvector of $M_{-\boldsymbol{k}}\mathcal{J}$ with eigenvalue $-\lambda_{\boldsymbol{k}}$. Hence, when $\omega_{\boldsymbol{k},n} \in \mathbb{R}$, $n=1,\dots,2m$ are the eigenvalues of $M_{\boldsymbol{k}}\mathcal{J}$, we can choose to set $\omega_{\boldsymbol{k}, n} = -\omega_{-\boldsymbol{k},n+m}$, where $\omega_{\boldsymbol{k}, n} > 0$ and $\omega_{-\boldsymbol{k},n+m} < 0$ for $n \leq m$ \cite{COLPA}. Then, $D_{\boldsymbol{k}}$ can be written $D_{\boldsymbol{k}} = \operatorname{diag}(\omega_{\boldsymbol{k},1}, \dots, \omega_{\boldsymbol{k},m}, \omega_{-\boldsymbol{k},1}, \dots, \omega_{-\boldsymbol{k},m})$, since $-\omega_{\boldsymbol{k},n+m} = \omega_{-\boldsymbol{k},n}$ [$D_{\boldsymbol{k}}\mathcal{J} = \operatorname{diag}(\omega_{\boldsymbol{k},1}, \dots, \omega_{\boldsymbol{k},2m})$].

\subsection{Energy correction}
When rewriting the Hamiltonian to the form in Eq.~\eqref{eq:H2matrix} we let
\begin{equation}
    \sum_{\boldsymbol{k}}a_{\boldsymbol{k}}^{(r)\dagger} a_{\boldsymbol{k}}^{(r)} = \frac{1}{2} \sum_{\boldsymbol{k}}(a_{\boldsymbol{k}}^{(r)\dagger} a_{\boldsymbol{k}}^{(r)} + a_{-\boldsymbol{k}}^{(r)} a_{-\boldsymbol{k}}^{(r)\dagger} -1).
\end{equation}
The commutator leads to an additional shift of the operator-independent part of the Hamiltonian,
\begin{equation}
    H'_0 = H_0 + \frac{N'}{2} \sum_r \pqty{2F_{\text{A},rr}^{(r)} - \eta_r} = H_0 + \frac{N'}{2} \sum_r \pqty{\nu_r - \eta_r}.
\end{equation}
The expectation value of the Hamiltonian is
\begin{align}
    \langle H \rangle =& H_0 + (H'_0-H_0) + \langle H_2 \rangle \nonumber \\
    =& H_0 + \frac{N'}{2} \sum_r \pqty{\nu_r - \eta_r} + \frac{1}{2}\sum_{\boldsymbol{k}}\sum_{n = 1}^{15} E_{\boldsymbol{k}, n},
\end{align}
at zero temperature. In SkX1, we find
\begin{align}
    H_0/NJ \approx& -7.530, \\
    (H'_0-H_0)/NJ \approx& -6.490, \\
    \langle H_2 \rangle/NJ \approx& 6.295, \\
    \langle H \rangle/NJ \approx& -7.725, \\
    (\langle H \rangle-H_0)/NJ \approx& -0.195.
\end{align}
The parameters are $D/J = 2.16$, $U/J = 0.35$, $K/J = 0.1$, $S = 1$ and we used 40000 points in the sum over $\boldsymbol{k}$. Notice that $\langle H \rangle < H_0$ which shows that quantum fluctuations stabilize the SkX. This agrees with Refs.~\cite{RoldanMolina, Sotnikov}. The quantum state is energetically preferred over the classical GS. Along with their small size, this is our justification for referring to the skyrmions in SkX1 and SkX2 as quantum skyrmions \cite{QSkOP, Sotnikov, Lohani}.

\section{Details of Chern number calculation} \label{sec:Cherndetail}
As shown in Fig.~2(c) in the main text, the Berry curvature develops strong peaks or valleys at values of $\boldsymbol{k}$ where the band has closely avoided crossings with other bands. This presents a numerical challenge in calculating the integral of the Berry curvature over the 1BZ. This is especially a challenge in SkX2, where many bands have closely avoided crossings at all values of $K/J$. We used the recursive algorithm in Ref.~\cite{AdaptQuad} to obtain adaptive quadratures, where the density of $\boldsymbol{k}$ values is largest around the sharp peaks and valleys. In general, this gave accurate numerical results, with deviations from integer Chern numbers decreasing to $\order{10^{-5}}$ or better with increasing number of $\boldsymbol{k}$ points in the quadrature. 

For $K/J = 0.71$ in SkX2, $E_{\boldsymbol{k},3}$ and $E_{\boldsymbol{k},4}$ are very close to crossing, with a gap of $\order{10^{-7}J}$. This yields extremely sharp peaks in the Berry curvatures and numerical difficulties led to $C_3 \approx 0.995$ and $C_4 \approx -0.995$ for both $10^6$ and $10^7$ $\boldsymbol{k}$ points in the adaptive quadrature. These are the greatest deviations from integers in our results. We view this as a numerical artifact since $K/J = 0.71$ is very close to the value $K = K_7$ where $E_{\boldsymbol{k},3}$ and $E_{\boldsymbol{k},4}$ cross. The numerical Chern numbers are better approximations of integers at values of $K$ farther away from $K = K_7$ than $K/J = 0.71$.

The derivatives in the Berry curvature in Eq.~(6) in the main text were calculated using forward difference with $\Delta k_{\mu} = 10^{-8}$. In the case of an evenly spaced discretization, the integral is converted to a sum via $\int d \boldsymbol{k} = (A_{\text{1BZ}}/N') \sum_{\boldsymbol{k}}$, where $A_{\text{1BZ}}$ is the area of the 1BZ, and $N'$ is the number of magnetic unit cells, i.e., the number of $\boldsymbol{k}$ points in the sum. 

The adaptive quadrature relies on subdivisions of the integration interval whenever the difference between two Gaussian quadratures of degree five and eight exceed a chosen tolerance \cite{AdaptQuad}. The application of Gaussian quadratures involve a change of variables in each subdivided integration interval and a generalization to a two-dimensional (2D) integral. The change of interval since Gaussian quadratures are designed for the interval $[-1, 1]$ can be performed as
\begin{align}
    \int_a^b f(x) dx &= \frac{b-a}{2} \int_{-1}^1 dx' f\left(\frac{b-a}{2}x' + \frac{a+b}{2}\right) \\
    &= \frac{b-a}{2} \sum_{i=1}^n w'_i f\left(\frac{b-a}{2}x'_i + \frac{a+b}{2}\right) \\
    &= \sum_{i=1}^n w_i f(x_i),
\end{align}
with $w_i = \frac{b-a}{2} w'_i$, $x_i = \frac{b-a}{2}x'_i + \frac{a+b}{2}$. The weights $w'_i$ and points $x'_i$ are provided by a Gaussian quadrature of degree $n$. The extension of Gaussian quadratures to 2D is given by 
\begin{equation}
    \int_{-1}^1 \int_{-1}^1 f(x, y) dx dy = \sum_i \sum_j w_i w_j f(x_i, y_j).
\end{equation}


\end{document}